%% file: ms.tex
\documentclass[12pt,preprint,natbib]{emulateapj}
\def \HST{{\emph{HST}}}
\usepackage{rotating}

\begin{document}
\slugcomment{1/25/10}

\title{The Asymptotic Giant Branch and the Tip of the Red Giant Branch as Probes of Star Formation History: The Nearby Dwarf Irregular Galaxy KKH~98}

\author{J. Melbourne \altaffilmark{1}, B. Williams\altaffilmark{2}, J. Dalcanton \altaffilmark{2}, S. M. Ammons\altaffilmark{3}, C. Max\altaffilmark{3},  D. C. Koo \altaffilmark{3}, Leo Girardi\altaffilmark{4}, A. Dolphin\altaffilmark{5} }

\altaffiltext{1}{Caltech Optical Observatories, Division of Physics, Mathematics and Astronomy, Mail Stop 301-17, California Institute of Technology, Pasadena, CA 91125, USA; jmel@caltech.edu}
\altaffiltext{2}{Department of Astronomy, Box 351580, University of Washington, Seattle, WA 98195, USA; ben@astro.washington.edu, jd@astro.washington.edu }
\altaffiltext{3}{University of California Observatories/Lick Observatory, Department of Astronomy and Astrophysics, University of California at Santa Cruz, 1156 High Street, Santa Cruz, CA 95064, USA;  ammons, max, koo@ucolick.org}
\altaffiltext{4}{Osservatorio Astronomico di Padova---INAF, Padova, Italy; leo.girardi@oapd.inaf.it}
\altaffiltext{5}{Raytheon, 1151 E. Hermans Road, Tucson, AZ 85706, USA; adolphin@raytheon.com}

\begin{abstract}
We investigate the utility of the asymptotic giant branch (AGB) and the red giant branch (RGB) as  probes of the star formation history (SFH) of the nearby ($D=2.5$ Mpc) dwarf irregular galaxy, KKH~98.  Near-infrared (IR) Keck Laser Guide Star Adaptive Optics (AO)  images resolve 592 IR bright stars reaching over 1 magnitude below the Tip of the Red Giant Branch.  Significantly deeper optical (F475W and F814W) Hubble Space Telescope images of the same field contain over 2500 stars, reaching to the Red Clump and the Main Sequence turn-off for 0.5 Gyr old populations.   Compared to the optical color magnitude diagram (CMD), the near-IR CMD shows significantly tighter AGB sequences, providing a good probe of the intermediate age (0.5 - 5 Gyr) populations.   We match observed CMDs with stellar evolution models to recover the SFH of KKH~98.  On average, the galaxy has experienced relatively constant low-level star formation ($5\times10^{-4}$ M$_{\odot}$ yr$^{-1}$) for much of cosmic time.   Except for the youngest main sequence populations (age $<$ 0.1 Gyr), which are typically fainter than the AO data flux limit, the SFH estimated from the the 592 IR bright stars is a reasonable match to that  derived from the much larger optical data set.  Differences between the optical and IR derived SFHs for 0.1 -- 1 Gyr populations suggest that current stellar evolution models may be over-producing the AGB by as much as a factor of three in this galaxy.  At the depth of the AO data, the IR luminous stars are not crowded.  Therefore these techniques can potentially be used to determine the stellar populations of galaxies at significantly further distances.  
\end{abstract}
  

\section{Introduction}
Star formation histories (SFHs) are a key measurement for understanding galaxy evolution in a $\Lambda$ Cold Dark Matter universe.  Two main approaches have been used to constrain the SFHs of galaxies.  One is to track the properties of different galaxy classes over cosmic time, such as star formation rate as a function of stellar mass \citep[e.g.][]{Noeske07}.  The other approach is to observe galaxies in the nearby universe and estimate their SFHs  based on their current properties \citep[e.g.][]{Heavens04}.  Results from both approaches suggest a ``downsizing'' in star formation  activity whereby the most massive galaxies form the bulk of their stars early (before $z\sim1$), while less massive, late type galaxies form the majority of their stars later \citep[e.g.][]{Brinchmann00}.  \citet{Noeske07} suggests that both the onset and duration of star formation may be correlated with mass, so called ``staged'' star  formation. For gas-rich dwarf systems, this scenario should result in relatively constant low-level star formation over much of a galaxy's lifetime. 
In this paper we derive the stellar populations for the local dwarf irregular galaxy KKH~98 as determined by optical/IR color-magnitude diagrams (CMDs) of the individual member stars.

\begin{figure*}
\center
\includegraphics[scale=0.8]{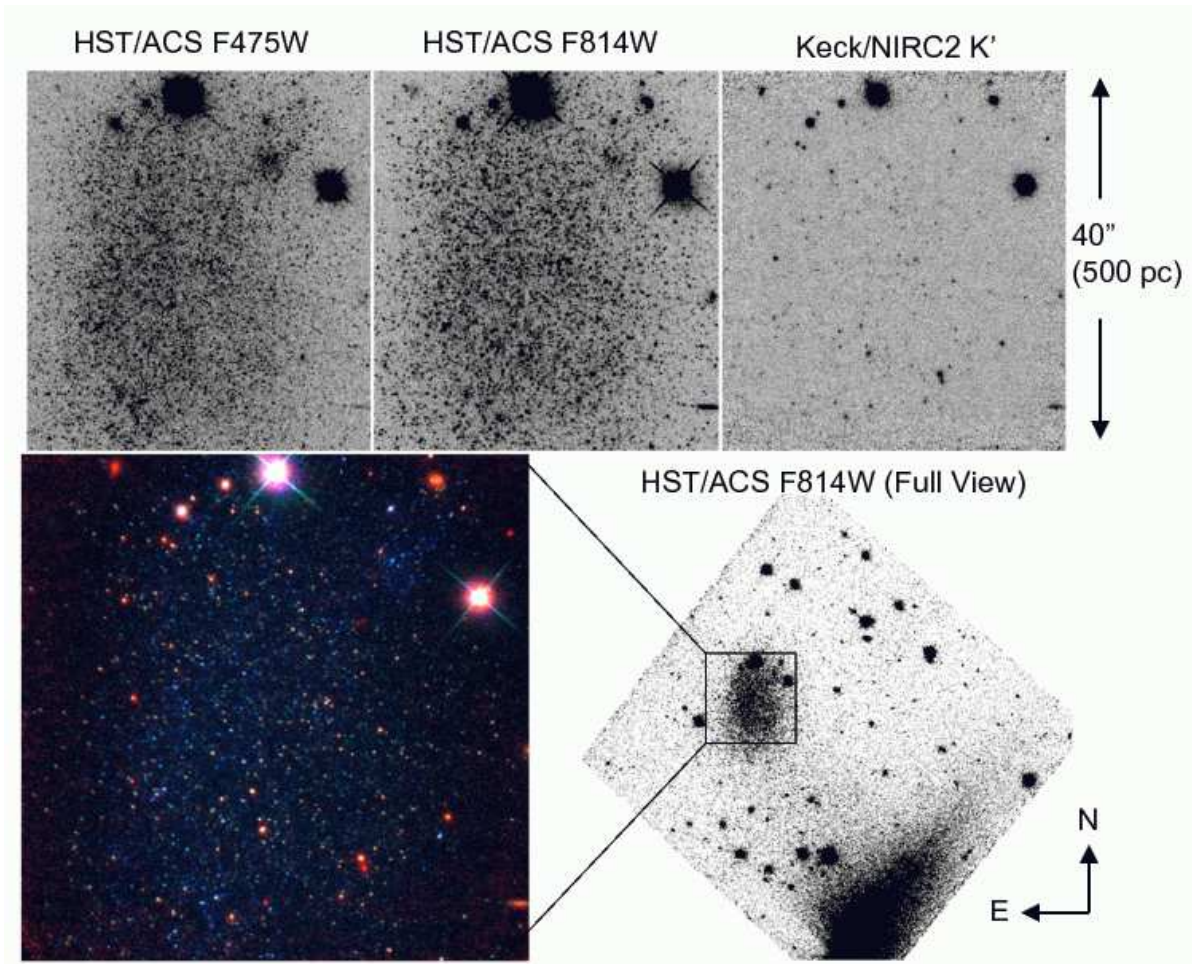}
\caption{\label{fig:KKH98} $40 \arcsec \times 40 \arcsec$ images of KKH~98 in the F475W, F814W, and $K'$-band (top row; left, middle, and right respectively).  Also shown are the  three color composite (bottom left), and the full ACS field of the \HST\ F814W-band image (bottom right).  North is up and East to the left in all images.  The spatial resolution of the Keck AO $K'$-band image is a good match to the resolution of the optical \HST\ images.  The F475W and F814W band \HST\ images reach depths of F475W$\sim27.6$ and F814W$\sim27.0$ (12\% photometric uncertainties).  Photometry of these images recovered 2539 stars within the AO field.  At this depth there is significant crowding in the \HST\ images.  In contrast, the Keck AO $K'$-band image probes the brightest 592 near-IR sources, and shows no significant crowding.  The very bright star at the top of the AO image was used as a tip-tilt guide star for the Keck AO system.}
\end{figure*}

For the most nearby galaxies ($d \la 4$ Mpc), color-magnitude diagrams of individual stars  
can be used to estimate SFHs.    Optical Hubble Space Telescope (\HST) imaging has been used extensively to study the stellar populations of nearby galaxies \citep[e.g.][]{Sarajedini05, Weisz08, Gogarten09, McQuinn09, Williams09a, Williams09b,Rejkuba09}.  For instance, deep \HST\ and ground based imaging of Local Group dwarf galaxies has revealed a subset of dwarf spheroidals and ellipticals with extended star formation histories and star formation episodes as recent as 1 Gyr ago  \citep[e.g.][]{Aparicio01,Dolphin05}.   The ACS Nearby Galaxy Survey Treasury \citep[ANGST;][]{Dalcanton09} has expanded the Local Group efforts to include a large sample of more distant galaxies  out to a distance, $d<4$ Mpc.  Using the \HST\ Advanced Camera for Surveys (ACS) and archival \HST\ imaging, ANGST has released photometry for deep two-color optical images of nearly 70 nearby galaxies. Matching the CMDs of these galaxies to model CMDs allows reconstruction of the star formation histories of a wide variety of galaxies in both group and field environments.     



Significantly less has been done to reconstruct the star formation histories from near-infrared (near-IR) stellar photometry.  Several programs have used the SFHs of local group galaxies, as measured from optical data, to test stellar evolution models in the near-IR.  For instance, \citet{Gullieuszik08} use the star formation history of Leo II dSph, obtained from optical CMDs, to estimate the expected production of AGB stars.  They find that the models over-predict the observed numbers of AGB stars by as much as a factor of six.  Likewise, \citet{Gullieuszik07} demonstrate how the near-IR colors of red giant branch (RGB) stars can be used to estimate the metallicity distribution within a galaxy.    Other studies have used the near-IR luminosity functions of the AGB and RGB populations to estimate SFHs of local group galaxies \citep{Olsen06,Davidge09}

In spite of the limited analysis to date, the near-IR has several features that make it attractive for probing SFH.  First, for most galaxies the bulk of the stellar luminosity is emitted in the near-IR \citep{Sawicki02}.  Second, intermediate aged asymptotic giant branch (AGB) populations are both extremely luminous in the IR \citep{Aaronson85, Frogel90, Maraston06}, and lie in tight sequences in near-IR color-magnitude space.  Third, the effects of reddening from dust are significantly reduced at near-IR wavelengths.  Fourth, RGB age/metallicity degeneracies are reduced for optical-IR colors compared to optical or IR data alone \citep{Gullieuszik07}.  Fifth, future missions such as the James Webb Space Telescope (JWST) and adaptive optics on the  Thirty Meter Telescope (TMT) will be IR optimized.  These missions will potentially resolve individual stars in galaxies out to the Virgo Cluster (18 Mpc) and beyond.  In order to take full advantage of these missions a clear understanding of the IR properties of the CMD are required.

 To test the efficacy of near-IR photometry for reconstructing the SFHs of nearby galaxies, we have begun a campaign to image ANGST galaxies  with Keck Adaptive Optics (AO) and \HST\ WFC3.  In this paper, we present results from Keck AO imaging of nearby dwarf irregular galaxy KKH~98. This galaxy was chosen because of its extended SFH and suitability of nearby AO guide stars.  We obtained $K'$-band (2.12 $\mu$m) Keck laser guide star (LGS) AO  imaging of  KKH~98, as part of the Center for Adaptive Optics Treasury Survey \citep[CATS][]{Melbourne05a,Melbourne08a}.  The spatial resolution of the AO image ($\sim0.1 \arcsec$) matched the spatial resolution of the \HST\ F475W and F814W images from ANGST.  592 individual IR bright stars ($K<23.5$) were resolved in the $K$-band, while 2539 stars were resolved in the optical (F475W $ <27.6$). We used the photometry of these stars to estimate the distance, metallicity, and star formation history of the galaxy.   

This effort is the first to use resolved IR photometry of individual stars to reconstruct the  complete star formation history (SFH) of a galaxy beyond the Local Group. We  compare the IR results to the well constrained, optically derived SFH of KKH~98, providing an excellent opportunity to test these techniques at longer wavelengths.  By refining SFH recovery methods in the IR, this work will help prepare for programs with JWST and TMT on more distant galaxies.

Throughout, we report Vega magnitudes and assume a $\Lambda$ cold-dark-matter cosmology: a flat universe with $H_0= 70$ km/s/Mpc, $\Omega_m=0.3$, and $\Omega_{\Lambda}=0.7$.


\section{\HST\ and Keck AO Observations of Dwarf Irregular Galaxy, KKH~98}
This study uses high-spatial resolution imaging data from both \HST\ and Keck AO to estimate the stellar populations within the dwarf Irr galaxy KKH~98.  KKH~98 was first identified by \citet{Karachentsev01} as a low surface brightness galaxy in the Palomar Observatory Sky Survey II plates.  It is a gas-rich, isolated dwarf irregular galaxy with an HI line width of $W_{50}=26$ km/s \citep{Karachentsev01}.    \citet{Karachentsev02} find a distance of 2.45 Mpc based on the tip of the red giant branch.  This distance gives an absolute magnitude of $M_B=-10.78$.  Its properties are fairly typical of the local dwarf irregular population identified by  \citet{Karachentsev01}.

\subsection{\HST\ Observations}
\HST\ observations of KKH~98 were obtained in both the F475W and F814W  filters of ACS as part of the ANGST survey \citep[see][for the details]{Dalcanton09}.   Three observations of the galaxy were made in each filter.  The combined images have an effective exposure time  of 2265 seconds and 2280 seconds, respectively, and are shown in Figure \ref{fig:KKH98}.  The ACS field of view ($202\arcsec \times 202\arcsec$) encompasses the bulk of the galaxy (See Figure \ref{fig:KKH98}). 
 
\subsection{Keck AO Observations}

On the night of November 24, 2007, we obtained $K'$-band (2.12 $\mu$m) Keck LGS AO    imaging of KKH~98.  We used the wide-field ($40\arcsec \times 40\arcsec$) camera of the NIRC2 instrument to achieve the largest field of view.  With a pixel scale of $0.04 \arcsec$/pix,  the wide-field camera under-samples the AO PSF ($0.06 \arcsec$ resolution), limiting the effective resolution to $\sim0.1 \arcsec$.  However, as is discussed below, the loss of resolution did not affect our ability to measure the photometry of the uncrowded IR-bright stars in KKH~98. 

Individual exposures were one minute, with a dither applied after two successive exposures.    During the observations, the laser was fixed to the center of the field.  Thus, as the telescope dithered, the laser dithered as well.  This method was preferred over fixing the laser to  the sky because of significantly lower overhead.  In addition, as was demonstrated in \citet{Steinbring08}, this method provides a more uniform AO correction over the field.  An $R=14.2$ tip/tilt star was located $\sim20 \arcsec$ north of the galaxy center (Figure \ref{fig:KKH98}).  The AO system used this star to track and correct for image motion, while the laser spot was used for higher order wavefront corrections.

We followed the image reduction method outlined previously in \citet{Melbourne05a}.  Sky and flatfield frames were created from the individual science frames after masking sources.  Frames were then flatfielded and sky subtracted.  We corrected frames for known NIRC2 camera distortions.  We aligned images by centroiding on sources, and combined them with a clipped mean algorithm.  The final reduced image has an effective exposure time of 45 minutes. Figure \ref{fig:KKH98} shows the resulting AO image of KKH~98.  It covers the central $40 \arcsec \times 40 \arcsec$ region of the galaxy.  

We observed the UKIRT photometric standard star FS3 immediately after the galaxy observations.  However, the night was not photometric.   Therefore we set the zeropoint for the AO data based on two bright unsaturated stars in the actual science frame which have $Ks$-band magnitudes from 2MASS \citep{Skrutskie06}.  Ultimately, we compared our photometry to stellar isochrones produced in the $K$-band system \citep{Bessell88}.  However  the transformation to the $K$-band system is negligible, typically $\la0.02$ mag for red stars (including the transformation from $Ks$ to $K$ \citep{Carpenter01} and the color terms to transform from $K'$ to $Ks$ \citep{Wainscoat92}).  Therefore, for simplicity, we heretofore report all AO magnitudes in the $K$-band Vega system.

The standard star observations of FS3 were obtained with tip-tilt AO correction to stabilize image motion, but without higher order LGS AO correction from the laser spot.  This strategy produced a high signal-to-noise ratio (S/N) observation of the profile of the wings of the AO PSF (which are primarily set by seeing and tip-tilt).  We used these standard star observations in the reconstruction of the AO PSF as outlined in the following section.  

\begin{figure}
\center
\includegraphics[trim=20 0 0 0,scale=0.6]{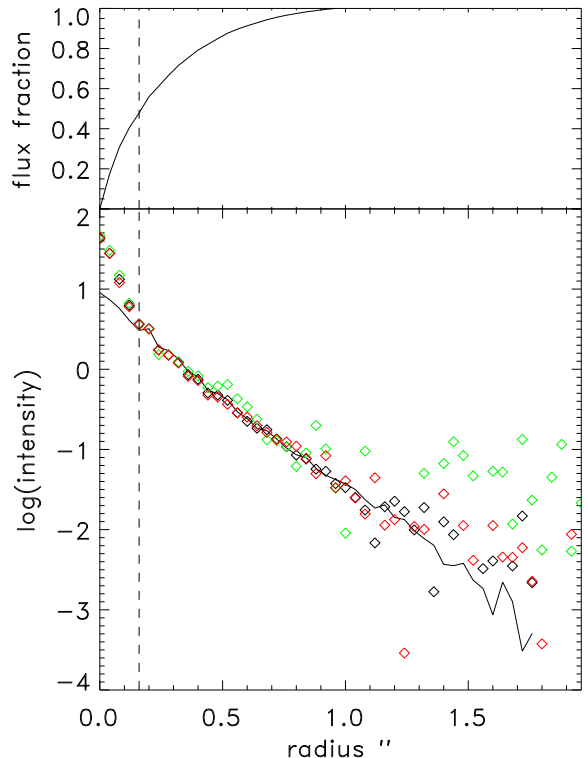}
\caption{\label{fig:PSFsplice} Bottom:  the azimuthally averaged stellar intensity as a function of radius (scaled to unity at a radius of $0.4\arcsec$) for the standard star FS 3 (solid line), and 3 stars in the AO image (black, green, and red diamonds).  FS3 was observed with tip-tilt correction on but no higher order AO correction.  The shape of this profile is therefore set by the seeing and tip-tilt correction.  This figure shows that for radii larger than $\sim0.2 \arcsec$, the halo of the AO corrected PSF has a similar profile as the tip-tilt corrected standard star PSF.  We therefore can use the high S/N PSF halo profile from our standard star (FS3) to track the profile of the our AO PSF out to large radii.The profiles of the 3 stars from the AO frame, which have higher order wavefront corrections, are significantly more peaked in their core than the standard star which only had tip-tilt correction.  Top: the fraction of light within a given radius for one of the AO corrected stars.  A vertical dashed line denoting a radius of $0.16\arcsec$ is shown.  This radius is the size of the aperture used in our photometry, and it contains more than 40\% of the stellar flux.  }
\end{figure}

\section{Photometry \label{sec:photometry}} 

\citet{Williams09a} and \citet{Dalcanton09}  describe the photometry  of the \HST\ ANGST images.  Stars in the ACS images were identified and measured with the  photometry routine DOLPHOT \citep{Dolphin00} using PSF fitting of individual sources with model \HST\ PSFs from TinyTim\footnote{http://www.stsci.edu/software/tinytim/}. Aperture corrections were measured for uncrowded stars and then applied to PSF fitting results.  Photometry lists were culled to remove extended objects and low S/N stars. Stars with S/N $>6$ in both filters were included in the final catalogue.  The matched F475W and F814W-band catalogue contains 11324 stars. Twelve percent photometric uncertainties were measured at F475W$=27.6$ and F814W$=27.0$.  Cutting the catalogue at these flux levels, and limiting to the field of view of the AO image results in a final set of 2539 stars.

The Keck AO image is significantly shallower than the \HST\ frames (Figure \ref{fig:KKH98}). However, the very deep \HST\ images already pinpoint the locations of stars in KKH~98.  We take advantage of the \HST\ positions when measuring stars in the AO image, performing photometry on the sources already identified in the \HST\  frames.  The AO PSF varies significantly across the field; thus we choose to use aperture photometry with aperture corrections  calculated across a grid of sub-regions in the image.  Discussions of the AO photometry --- including the PSF, aperture corrections, and aperture photometry --- are given below.


\begin{figure}
\includegraphics[trim=20 0 0 0,scale=0.6]{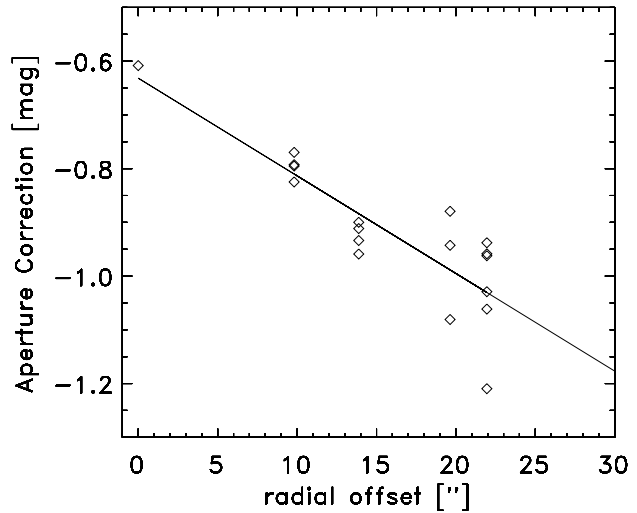}
\caption{\label{fig:apcor}  The aperture correction for an $0.16\arcsec$ aperture as a function of separation from the center of the AO image.  The correction increases linearly with distance to the image center.  We fit all data points with a line, giving a unique correction for each star in the image.  This fit also improves the aperture correction for stars on the outskirts of the image where there are few stars to make a good estimate of the PSF.
}
\end{figure}

\subsection{AO PSF}
To perform accurate photometry, we need a model of the AO PSF.  Unfortunately, the AO correction degrades spatially with distance from the laser spot (anisoplantism) and the tip-tilt star (anisokinetisism), producing a spatially varying PSF.  We model  the PSF using two components, (1) an AO corrected core with a FWHM of roughly the diffraction limit of the telescope, and (2) a halo with a profile width set by the atmospheric seeing and AO tip-tilt correction. While the shape of the PSF halo is relatively stable across the image, the core varies significantly with position.  

To get an accurate representation of the profile of the PSF core, we identify luminous, isolated stars within 25 equal-area sub-regions of the image. After subtracting neighbors, we register and coadd the stars in each sub-region, and trace the PSF profile of the coadded image. 

Unfortunately, the PSF halo is spread over hundreds of pixels, and is much lower S/N than the diffraction limited core. Therefore its profile cannot be accurately traced even for the stacked PSFs described above.  Instead we used images of the very bright flux standard star, FS 3, observed with tip-tilt AO correction on, and  high order AO correction off.  With high order correction off, light that would have been corrected into the PSF core by the AO system, remained in the halo. The resulting PSF has a very high S/N profile out to large distance ($\sim2\arcsec$) from the PSF center, and is a good match to the halo of the AO PSF (Figure \ref{fig:PSFsplice}).   

  
To produce a final PSF for each of the 25 sub-regions, we splice together the halo and core images, scaling the halo image to map smoothly onto the image of the core.  As was done in \citet{Melbourne08a}, we select the splice point to be the radius at which the flux in the azimuthally averaged core profile is 9 sigma above the sky noise, as this is a good match to bright AO corrected stars where the halo is easily observed.  Figure \ref{fig:PSFsplice} shows that the PSF profile is relatively independent of the particular choice of splice point, for splice points larger than several times the diffraction limit (e.g. $\sim 0.2 \arcsec$).   This figure also shows that the halo PSF profile from the standard star is a good match to the few AO corrected stars in the science frame that are bright enough to adequately detect the halo.

\subsection{AO Aperture Corrections}
The stellar photometry of the AO image was performed with small apertures on the high S/N core of the AO PSF.   We chose a 4 pixel radius aperture ($0.16 \arcsec$). This aperture is 50\% larger than the typical FWHM of the stellar PSF, and maximizes the S/N of the aperture photometry.  We convert the small aperture photometry to total flux with aperture corrections calculated from the 25 model PSFs, one for each sub-region of the image.   Total flux for the model PSFs was measured with a curve of growth technique out to $2\arcsec$.  

The aperture correction varies from 0.6 mags at the center of the field to $\sim1$ magnitude near the edge, where a larger fraction of the light remains uncorrected. Figure \ref{fig:apcor} shows the variation of the aperture correction across the field.  A linear fit to the data gives aperture correction as a function of distance from the image center for an arbitrary stellar position.   The fit provides a more finely sampled measure of the aperture correction across the field and improves the estimates in the outskirts of the image where the PSF is poorly measured due to a lack of stars.   We use this functional form for the aperture corrections in the final photometry.

\begin{figure}
\center
\includegraphics[trim=20 0 0 0, scale=0.6]{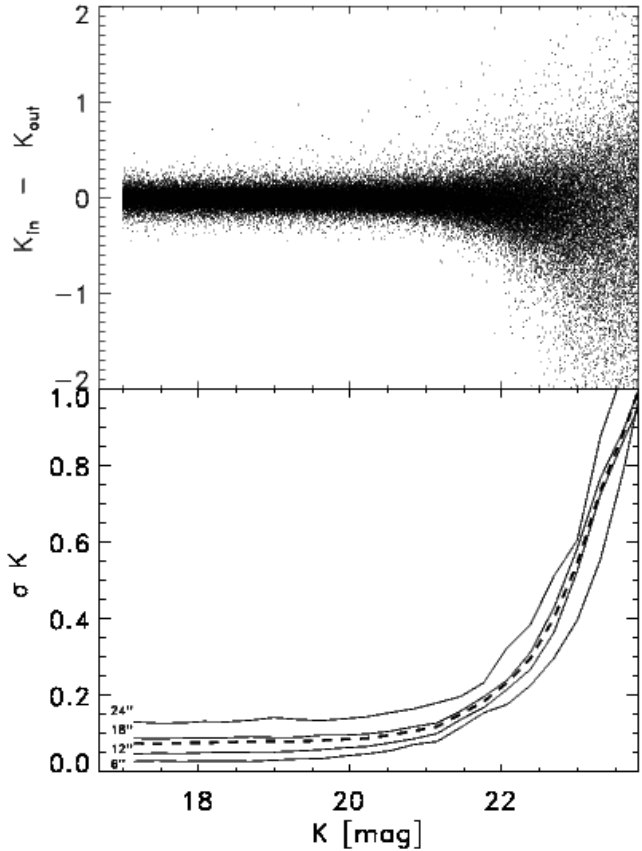}
\caption{\label{fig:model_phot} Photometry of 500,000 artificial stars embedded at random locations in the AO image.  The top panel shows the difference of input vs. measured magnitude while the bottom panel shows the standard deviation of the difference as a function of input magnitude.  The thick dashed line represents the standard deviation calculated from all of the stars.  The thin lines represent the standard deviations of groups of stars with consecutively larger separations from the image center (6, 12, 18, and 24 arcsec from the image center, as marked).  Because the PSF is tighter near the center of the AO image, the S/N is higher for the photometry done there.  The photometry shows low photometric uncertainty (better than 15\%) for stars brighter than $K=22$.  }  
\end{figure}

\subsection{AO Photometry \label{sec:phot}}
The deep \HST\ images of KKH~98 identify the locations of stars in the galaxy.    To translate the \HST\ positions onto the Keck AO image, we first selected a set of 157 bright stars that are easily visible in both the \HST\ and AO images.  We then used the IRAF CCTRAN program to  transform the \HST\ coordinates into the AO image coordinates, using.  a quadratic transformation.  Since the geometric distortions in the AO image were removed during image processing, this transformation worked well across the entire frame.

We measured the $K$-band magnitude for each star in the \HST\ star list that was located on the AO frame.  For each star, bright neighbors were identified with Sextractor.   Scaled model PSFs were used to subtract the bright neighbors using a PSF fitting technique.  We then  measured aperture photometry on the star of interest, applying the aperture correction appropriate for the radial separation of the star from the center of the image (Figure \ref{fig:apcor}).  A correction for the local sky was made, using a median sky  value measured in an annulus of  radius of $1.0 -1.2$ arcsec.  When making the sky measurement, stars identified in the sky annulus  by Sextractor segmentation maps and the \HST\ star lists were masked out. 

Two types of uncertainty dominate the error budget in these measurements, (1) the photometric uncertainty, set primarily by the Poisson sky noise, and (2) the uncertainty introduced by the spatially varying PSF and aperture corrections.   The photometric Poisson uncertainty can be measured directly from the images.  For each star, we measure the standard deviation of pixels in the nearby sky to estimate this uncertainty.  The uncertainty introduced by the aperture corrections can be measured from the scatter  in Figure \ref{fig:apcor}.  While the scatter is small near the center of the image, it increases significantly towards the edges of the image.  

To estimate any systematics in the measurement methods, we also performed a Monte Carlo simulation, embedding 500,000 artificial stars into the image at random locations with sub-pixel sampling.  Artificial stars ranged in magnitude from $K=17$ to 24, but were all assumed to be identified in the corresponding \HST\ image.  We measured the photometry of the artificial stars in the exact same manner as the real stars in the \HST\ catalogue.  Figure \ref{fig:model_phot} shows the standard deviation of the photometry as a function of input $K$-band magnitude.  We find to $K=22$, the photometry is better than 0.2 mags.  At $K=23.5$ the photometric uncertainty is about a magnitude.

An additional 22 IR-bright stars were identified in the AO image that were not included in the final \HST\ photometry lists, usually because they fell below a blending or S/N threshold in the F475W image.  We found that these stars do not alter the conclusions of this paper and therefore do not include them in the final analysis.
    
\begin{figure*} 
\centering
\includegraphics[trim=60 0 0 0, scale=0.8]{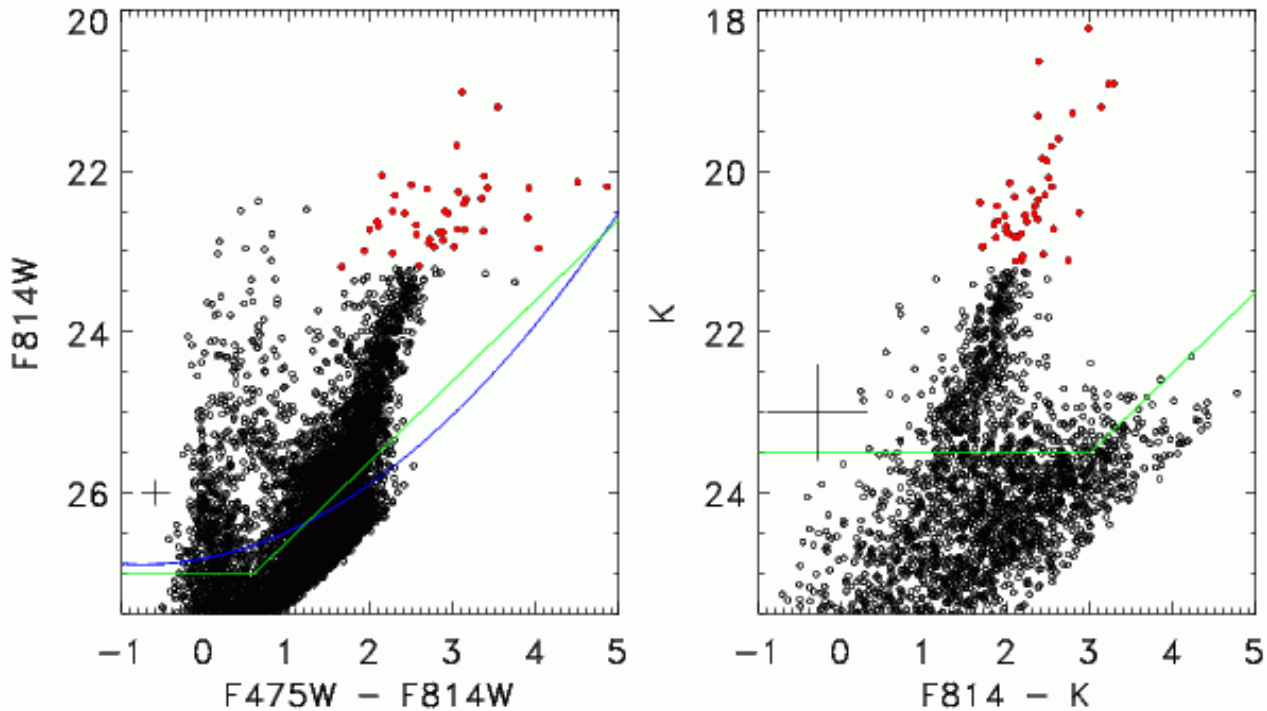} 
\caption{\label{fig:CMD}  Apparent (uncorrected for reddening) color-magnitude diagrams of stars in KKH~98.   F475W $-$ F814W color is plotted on left, and F814W$-K$ color is plotted on the right.  Suspected AGB stars are shown as filled red circles.  While both the optical and IR CMDs show a tight red giant branch, the IR color shows a much tighter AGB sequence, resulting in a powerful age indicator for intermediate aged populations.  The 50\% completeness limits (blue curve), and  the cuts applied to the data, for modeling purposes (green lines) are shown for the optical data.  The data cuts applied to the IR data are also shown. Typical photometric uncertainties at the faint end of each CMD are shown.  At the bright end the uncertainties are roughly the size of the data-points. Data at the faint end of the IR CMD ($K<23$) have photometric uncertainties greater than 0.6 magnitudes.  Thus the large scatter in the IR CMD at faint magnitudes is not a feature of stellar evolution but rather photometric error.} 
\end{figure*} 

\section{Results}
Figure \ref{fig:CMD} shows the F475W - F814W vs. $F814W$ CMD and the F814W$-K$ vs. $K$ CMD for KKH~98.  Both CMDs show a tight red giant sequence with a tip at  F814W$=23.2\pm0.1$ and  $K\sim21.2 \pm 0.1$  (based on a visual fit to the CMD).   The optical CMD exhibits a significant  blue upper main sequence ($F475W - F814W\sim0.0$) suggesting the presence of very young stars in this galaxy.  The young red helium burning sequence is poorly populated, although there do appear to be some young blue core helium burning stars. The main sequence is not obvious in the F814W$-K$ CMD, because of the flux limits of the $K$-band data.  

In the IR CMD, a well-defined asymptotic giant branch (AGB) emerges from the tip of the red giant branch (TRGB) and extends to $K=18.5$.  Compared with the optical CMD, the AGB stars in the F814W$-K$ CMD form a much tighter sequence.  Tighter AGB sequences are expected at redder wavebands for several reasons: (1) lower amounts of self-extinction from dusty AGB envelopes, (2) a flatter color vs.\ stellar T$_{eff}$ relation \citep[e.g.][]{Gullieuszik07}, and (3) the thermally-pulsing AGB stars have much smaller pulsation amplitudes \citep{Whitelock06, Marigo08}.  Because the AGB produces tight sequences in near-IR color-magnitude space, the AGB should prove a powerful indicator of star formation rate for intermediate aged populations, i.e. older than $\sim0.1$ Gyr.

\begin{figure*} 
\centering
\includegraphics[trim=60 0 0 20, scale=0.75]{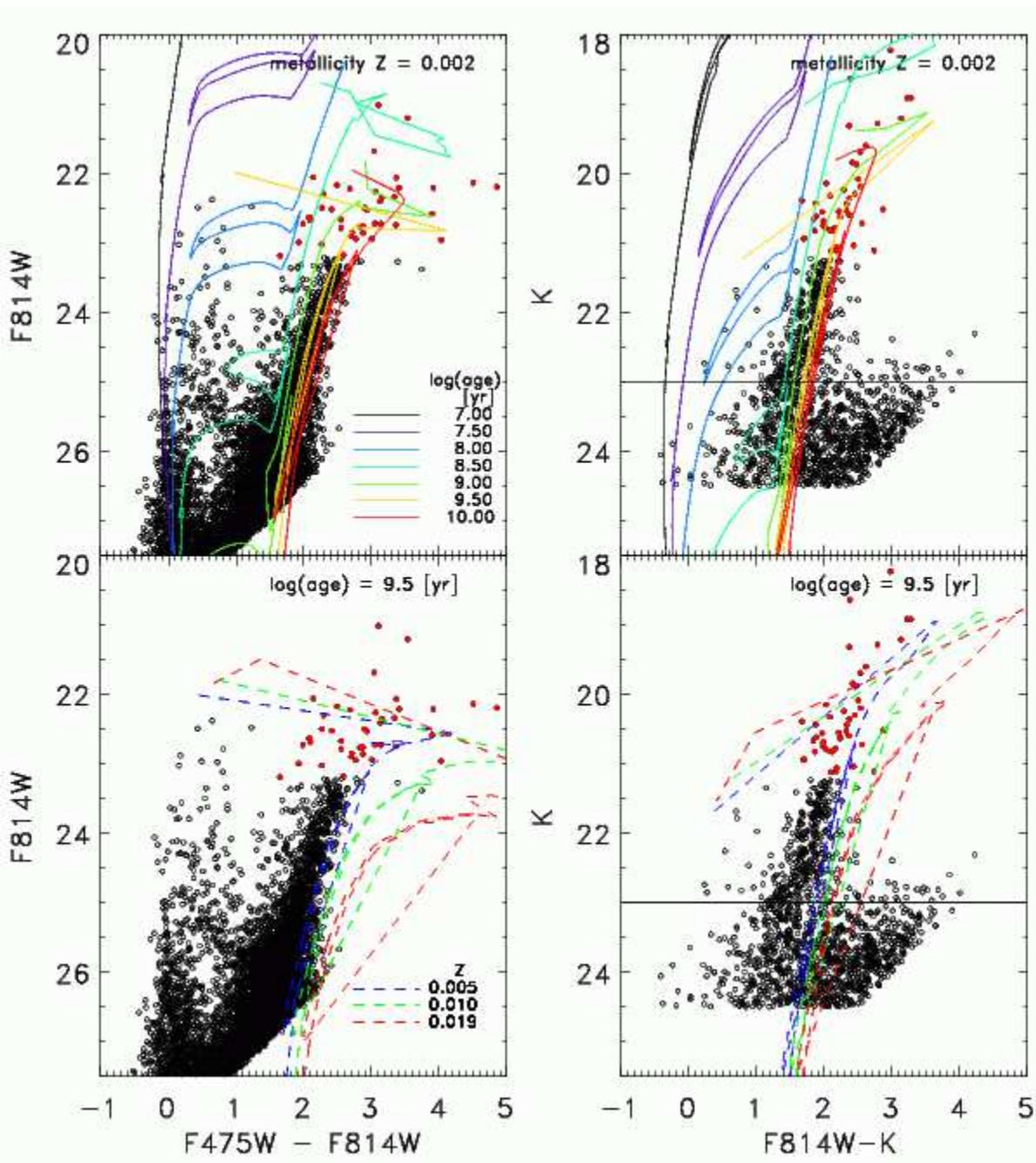} 
\caption{\label{fig:CMD2}  Same as Figure \ref{fig:CMD} only now with model isochrones overplotted \citep[from][]{Marigo08}. Top: metal poor isochrones, 1/10th solar, spanning ages from $10^7 - 10^{10}$ years (colored lines) nicely span the data. Bottom: more metal rich isochrones (dashed lines) with an age of $10^{9.5}$ years, the age by which roughly 80\% of the stars had formed, are also shown and are significantly redder than the data. This suggests that the stars of KKH~98 are metal poor. Note: the very red colors measured for some faint stars in the IR CMD (below the horizontal line) can be explained by photometric uncertainty (greater than 1 mag.), rather than some usual stellar population. } 
\end{figure*} 

At the faint end of the IR CMD (i.e. $K>23$), photometric uncertainties in the $K$-band produce significant scatter in the photometry.  Thus the large scatter to red colors in Figure \ref{fig:CMD} (and Figure \ref{fig:CMD2}) is not a physical feature in the data (such as the red clump), just photometric noise.  

Because of the spatially varying AO PSF, we investigated the possibility that the AO photometry might vary systematically across the image.  However, CMDs generated from multiple sub regions across the image show no systematic deviations.

\begin{figure}
\center
\includegraphics[trim=10 0 0 0,scale=0.5]{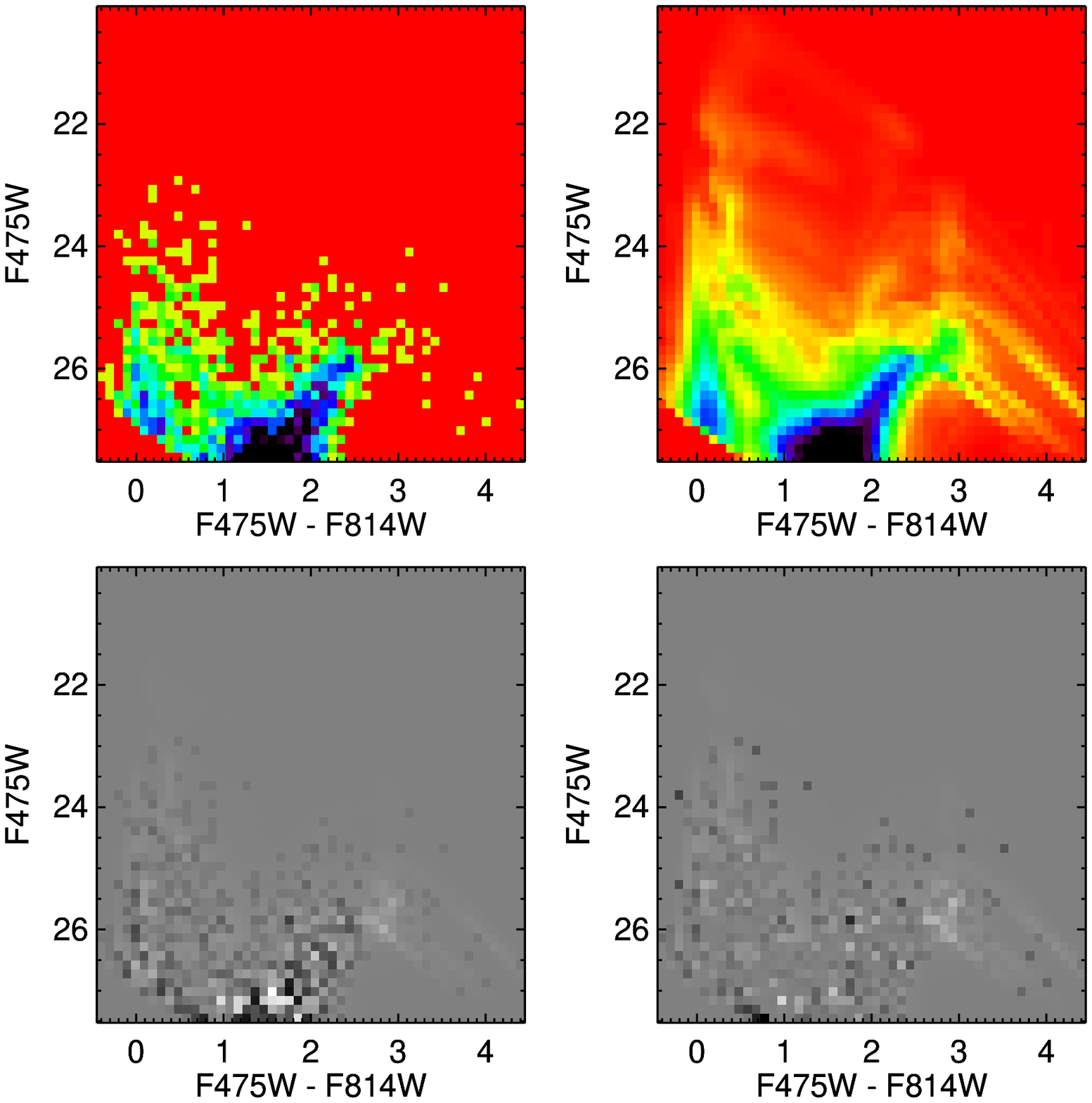}
\caption{\label{fig:CMDresid} TOP: the optical CMD now displayed as a density Hess Diagram.   The true data (top left), and the best-fit MATCH-derived model (top right) are shown.  Also shown are the residual differences between the two (bottom left), and the residual difference normalized by Poisson statistics in each bin (bottom right).  In the residual maps, darker regions denote an under-production of model stars. Lighter bins denote an over-production of model stars. The residuals (bottom left) vary from +10 stars per bin (white) to -11 stars per bin (black). The normalized residuals (bottom right) vary from +8 (white) to -5 (black) and can be thought of as sigma deviations per bin. The models reproduce the data, except at the faint end, and for the AGB, where the models over-produce AGB stars by roughly a factor of three (F475W $\sim25.7$ and F475W - F814W $\sim3$, see Figure \ref{fig:bi_fixbi} for more details).  }
\end{figure}

\begin{figure}
\center
\includegraphics[trim=10 0 0 0,scale=0.5]{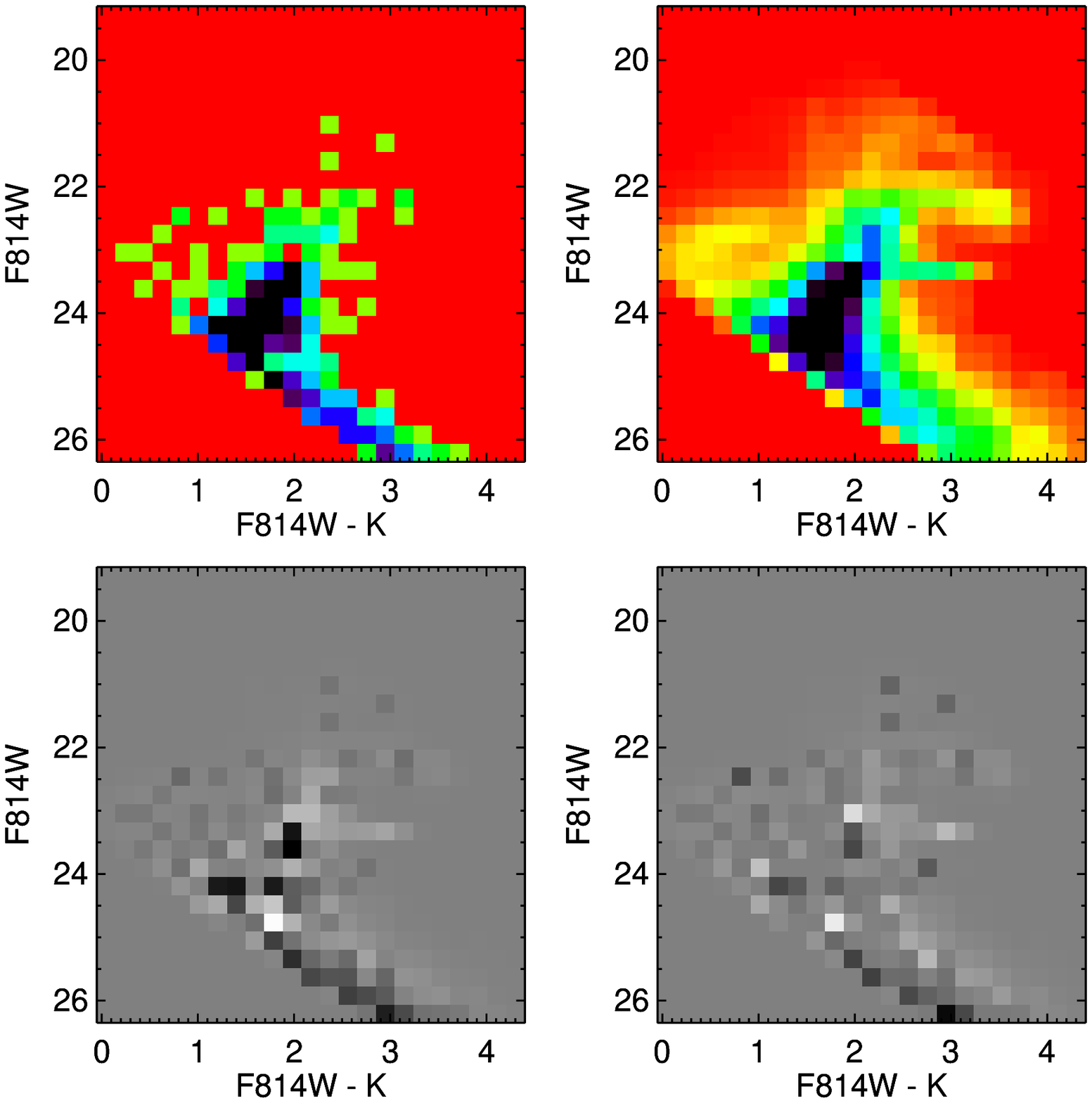}
\caption{\label{fig:CMDresid2} Same as Figure \ref{fig:CMDresid}, only now for the IR data.   The residuals (bottom left) vary from +10 (white) stars to -9 stars (black), while the normalized residuals vary from +5 sigma to -5 sigma. The fit to the AGB is better for the IR CMDs than the optical CMDs, although the models still over-predict the the AGB by roughly a factor of two (see Figure \ref{fig:ik_fixik} for details).      
}
\end{figure}

\subsection{Modeling the Color-Magnitude Diagrams \label{sec:sfh}}
We estimate the star formation history (SFH) of KKH~98 by comparing the observed CMDs with CMDs produced from the stellar isochrones of \citet{Girardi02}, with updated AGB models from \citet{Marigo08}.  These isochrones were used together with the bolometric corrections and color transformations from \citet{Girardi08}.  Sample isochrones, overlayed on the CMDs, are shown in Figure \ref{fig:CMD2}. Metal poor (1/10th solar) isochrones (top row Figure \ref{fig:CMD2}) span the photometric data.  In contrast, more metal rich isochrones (bottom row Figure \ref{fig:CMD2}) tend to be redder than the data, suggesting that the stars in KKH~98 are metal poor.  

 We use the MATCH package \citep{Dolphin02} to fit the CMD for a star formation history, metallicity history, dust reddening, and distance modulus (DM), for an assumed Salpeter IMF and binary fraction of 0.35.  MATCH uses a Poisson Maximum Likelihood statistic to determine the best fit model and the 1 sigma uncertainties on that model.  For a complete description of the MATCH routine, see \citet{Dolphin02}.  

\citet{Williams07} showed that the IMF choice does not affect the relative star formation rates (as a function of time), but can change the overall normalization.  Our primary goals are to compare the SFH measured from the optical data to the SFH measured from the IR data, and to determine if the SFH is ``bursty'' or continuous.  As these are relative measurements, neither will be significantly affected by choice of IMF or binary fraction.

Additional inputs to MATCH include: (1) a range of acceptable distances and reddening values; (2) time step sizes; and (3) ranges and binning sizes in color-magnitude space. In general we, attempted to reproduce the choices made in previous ANGST work \citep{Williams07, Williams09a}.  Distance ranges were allowed within $\pm0.5$ mags of the distance modulus in the literature of $DM=26.95 \pm0.11$ \citep{Karachentsev02}.  Reddening was allowed to vary from $A_V=0.1 - 0.7$, a reasonable range for a dwarf galaxy with foreground reddening $A_I=0.24$ as measured in the \citet{Schlegel98} Galactic dust maps.  We use a series of 12 logarithmically spaced time-steps which allow us to explore the star formation history in both the recent (Myrs) and distant (Gyrs) past.  The minimum sizes for color-magnitude bins are set by the uncertainties of the stellar isochrones and are roughly 0.05 mags in color and 0.1 mags in magnitude.  However, when setting bin sizes, additional considerations include allowing for reasonable number statistics in each bin while providing the resolution necessary to resolve features in the CMDs.  For the optical data we used color bins of size 0.1 mags and magnitude bins of size 0.15 mags.  We doubled the sizes of these bins in the NIR to offset the smaller number statistics in the NIR CMDs.  Small changes in the bin sizes were not found to alter the results.

We used the latest version of MATCH (version 2.3) which handles the isochrone color transformations for Carbon and Oxygen AGB stars independently.  In older versions of MATCH, C/O ratios were not tracked and the color transformations produced AGB stars significantly redder (2-3 mags) than expected.  The colour transformations of C stars in the updated version of MATCH were derived from \citet{Loidl01}. Because the information content of the IR CMD is biased towards the AGB populations, it was important for us to use the most realistic prescriptions for  AGB photometry.  In the optical CMDs these considerations were less vital because the information content at young and intermediate ages primarily arises from the Main Sequence turn-offs.

For the optical data, we chose to model the CMD with photometry brighter than the 12\% uncertainty limits and 50\% completeness limits (F475W$<27.6$ and F814W$<27.0$) to avoid fitting areas of the CMD with poor photometry or large completeness corrections. These limits result in photometry for 2539 stars in the area overlapping the AO data.  For the IR CMD, we chose  stars with $K < 23.5$, the point at which the $K$-band photometric uncertainty is about 1 magnitude.  These cuts provide 592 stars.  While the photometric precision at the fainter end of $K$ is poor, the long wavelength baseline between the F814W and $K$-bands means that there is still statistically interesting information at these fainter limits.  Table \ref{tab:results} gives a summary of the MATCH derived results which are discussed in detail below. 

Figures \ref{fig:CMDresid} and \ref{fig:CMDresid2} show Hess diagrams of the CMDs in the optical and IR (top left).  Also shown are the best fit model CMDs derived by MATCH (top right), the residual difference between the model and the data (bottom left), and the residuals scaled by the statistical significance of bin (bottom right).  The residual differences between the data and MATCH derived models are small for both the optical and IR CMDs, except for AGB stars, especially in the optical data where the models predict 130 AGB stars compared to 42 observed (with the caveat that the AGB must be brighter than the TRGB, and redder than $F475W - F814W = 1.75$). In the IR, the models predict 92 AGB stars (brighter than the TRGB).  We return to these discrepancies in the following section and elaborate in the discussion section. In contrast, the numbers of stars in other evolutionary stages, such as the red giant branch and main sequence are better reproduced by the models.  Table \ref{tab:cmdnum} gives the measured and predicted numbers of stars in different regions of the KKH~98 CMD. 
\begin{figure}
\center
\includegraphics[scale=0.4]{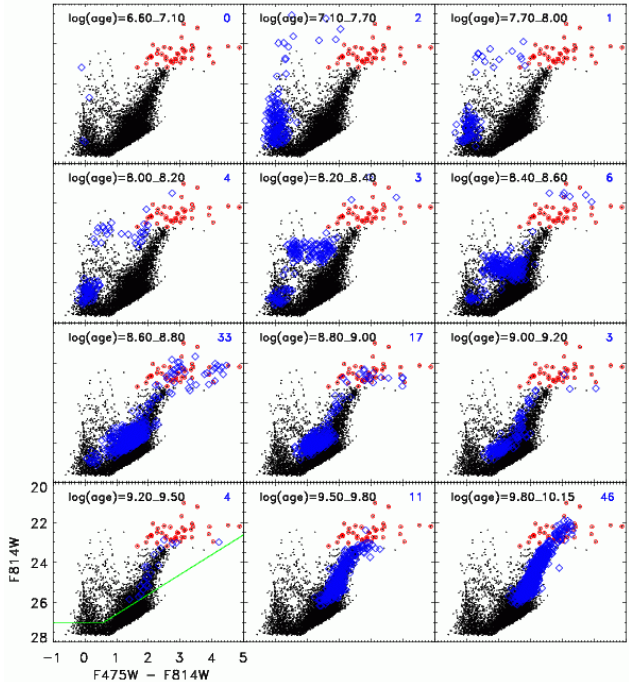}
\caption{\label{fig:bi_fixbi} A realization of the best fit optical model CMD shown broken into different time bins (blue diamonds).  The observational data are also shown (black points), with the AGB stars circled in red.  The models shown that the main sequence is populated by young stars (log(age) $< 8.8$), whereas the red giant branch is primarily populated by old stars (log(age) $>9.5$). The model predicted numbers of AGB stars are given in the upper right of each CMD. For AGB stars above the TRGB (and $F475W - F814W > 1.5$), the data show 42 stars, while the models predict 130 stars. The lower left plot shows the magnitude limits provided to MATCH (green lines). }
\end{figure}

\begin{figure}
\center
\includegraphics[scale=0.4]{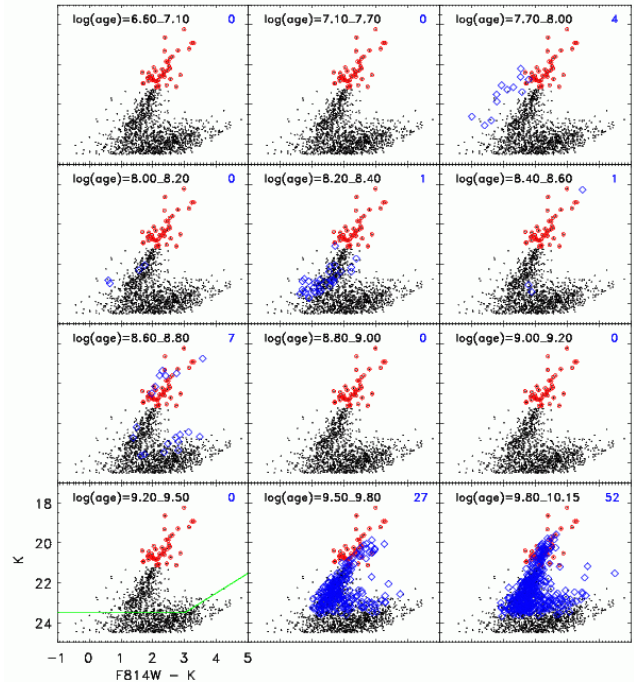}
\caption{\label{fig:ik_fixik} Same as Figure \ref{fig:bi_fixbi}, only now a realization of the best fit IR model CMD (blue diamonds) shown broken into different time bins.   The model predicted numbers of AGB stars are given in the upper right of each CMD. For AGB stars above the TRGB, the data show 42 stars, while the models predict 93 stars.  In this case we see that the bulk of the predicted AGB stars are old (log(age) $<9.5$). The lower left plot shows the magnitude limits provided to MATCH (green lines). }
\end{figure}

\subsubsection{Star Formation Histories}
\begin{figure}
\center
\includegraphics[trim=50 150 0 20,scale=0.6]{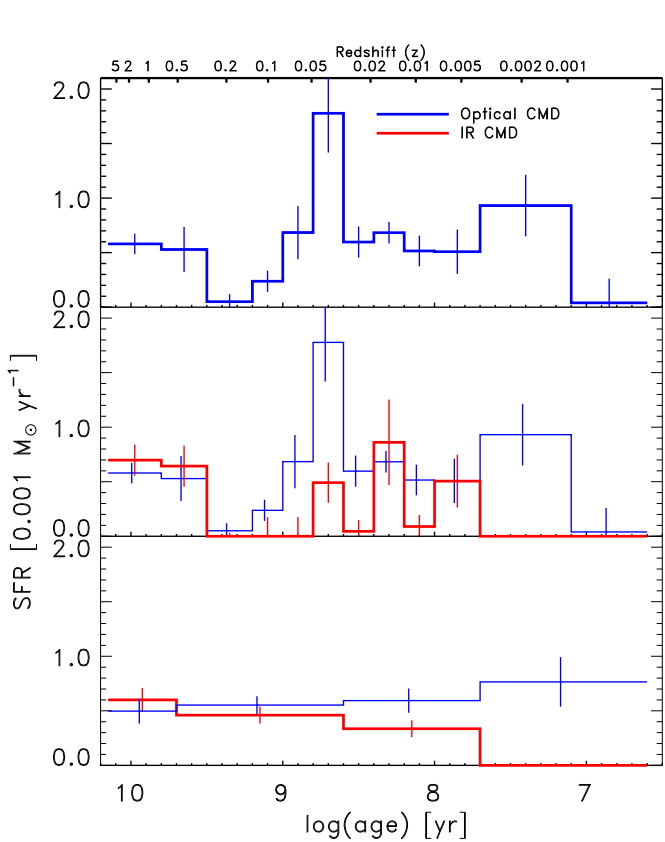}
\caption{\label{fig:sfh} The MATCH derived star formation history of KKH~98 from the optical CMD (blue) and the IR CMD (red). At early times, both show a relatively constant star formation rate of  $5\times10^{-4} M_{\odot}$ yr$^{-1}$, or $2 \times10^{-4} M_{\odot}$ yr$^{-1}$ kpc$^{-2}$.  From $1-3$ Gyrs ago, both show a significant drop in the star formation to negligible rates.  More recently, (age $< 1$ Gyr) the SFR has roughly returned to its past average.  However, the optical data suggest a possible burst of star formation at 0.5 Gyrs ago, not reproduced in the IR data.   The IR CMD is also insensitive to the most recent star formation (i.e. age $<$ 60 Myr).  Uncertainty estimates are shown for each time bin, and combine in quadrature (1) the systematic uncertainty between the models and the data, and (2) the uncertainty associated with small number statistics in a given time bin.  The bottom panel shows the data rebinned to longer time periods.               
}
\end{figure}

The optical and IR derived SFHs of KKH~98 are plotted in Figure \ref{fig:sfh}.  For the bulk of cosmic time, the SFRs from the two CMDs agree to within the uncertinties.  For ages older than 3 Gyrs, both data sets show evidence for a modest, roughly constant star formation rate of $5\times10^{-4} M_{\odot}$ yr$^{-1}$, or $2 \times10^{-4} M_{\odot}$ yr$^{-1}$ kpc$^{-2}$.  From $1 - 3$ Gyrs ago,  both show a drop in star formation to roughly negligible rates.  More recently, both show renewed star formation for ages $ < 0.5$ Gyrs ago.  However, in these youngest time bins, there are differences in the details. At the youngest ages (age $< 0.1$ Gyr), the optical photometry suggests ongoing star formation, while the IR data set does not. This difference is not alarming, as the near-IR data are not very sensitive to hot (blue) main sequence stars produced by recent star formation. 

Another difference is that the optical data set  suggests a burst of star formation, at age $\sim0.5$ Gyrs (Figure \ref{fig:sfh}).  In contrast, the IR data suggest SFRs roughly a factor of four smaller at that time.  The evidence for a burst in the optical data comes from stars at the well-populated main sequence turn off (MSTO) at the faint end of the optical CMD.  This is shown in Figure \ref{fig:bi_fixbi} where a realization of the predicted model CMD divided into different time bins.  At log(age) $= 8.6 - 8.8$ [yr], large numbers of stars are predicted at the MSTO.  In contrast the IR data do not reach deep into the main sequence, and therefore the IR-derived SFH is not sensitive to the main sequence turn-off feature.  However, the IR data should contain information on populations of that age from the AGB sequence (see Figure \ref{fig:ik_fixik}).   Therefore, if the burst is real, evidence for it should exist in the AGB population, assuming the AGB models are accurate.  In the discussion section we examine possible explanations for the measured differences in the SFR at these ages.  Here we discuss the effect of time binning.

Time binning may play a role in the differences between the optical and IR derived SFHs.  Logarithmic time bins have been selected to fully explore the SFH on both recent and older time scales.  However, especially for the IR data, which has only 592 stars, the time binning may be more fine than the data warrant. The third panel of Figure \ref{fig:sfh} shows the IR results now binned with larger time steps. These bins were selected by first binning the data on small time scales and then tracking the goodness of fit parameters as bins were dropped from the sample.  If the fits did not change by more than the measured uncertainties, a dropped bin would be incorporated with its neighboring bin.  Ultimately, this resulted in 4 bins.  We caution, however, that this method is likely to miss the importance of  bins with little star formation such as those from $1 - 3$ Gyrs ago. For more on optimal binning techniques, see Williams et al. (2009 in press). The optical data have been binned to match the IR bins.  Binned this way, there are still some differences between the two measured star formation histories at the youngest ages. 

The SFR uncertainties shown in Figure \ref{fig:sfh} are a combination of (1) the systematic uncertainty between the models and the KKH~98 photometry, and (2) the random uncertainty associated with small number statistics of stars in any given color-magnitude bin.  The first uncertainty is estimated by the MATCH routine from the maximum likelihood statistic as it attempts to  fit models parameters within the adopted ranges.  The second uncertainty is estimated by a Monte Carlo bootstrap simulation of 100 realizations, where we rerun MATCH but alter the input photometry list, so that it is composed of random draws from the actual photometry list.  The systematic uncertainties are roughly equivalent for both the optical and IR data, and are similar in scale to the random uncertainties associated by small number statistics.     The error bars shown in Figure \ref{fig:sfh} are the two uncertainties added in quadrature.  

Surprisingly, the optical and IR model uncertainties are similar in scale despite the fact that the optical data contain contain a factor of 5 more stars.  Several factors conspire to lead to this result.  First, the color-magnitude bins used in the IR analysis are twice as large as the optical analysis, and thus typically contain as many or more stars as the optical bins.  Second, a lack of stars in a feature can be just as important to producing a significant result as the presence of stars.  For example the lack of AGB stars in the data forces the IR models to select against a burst of star formation at 0.5 Gyrs ago (see Figure \ref{fig:ik_fixik}), resulting in small numbers of AGB stars in both the models and data, and a low uncertainty assigned to this time bin.  Whereas in the optical data, the main-sequence turn off suggests a burst at that time, and the lack of AGB stars acts to increase the uncertainty of the high SFR measured in this time bin.

\begin{figure}
\center
\includegraphics[trim=30 0 0 0,scale=0.55]{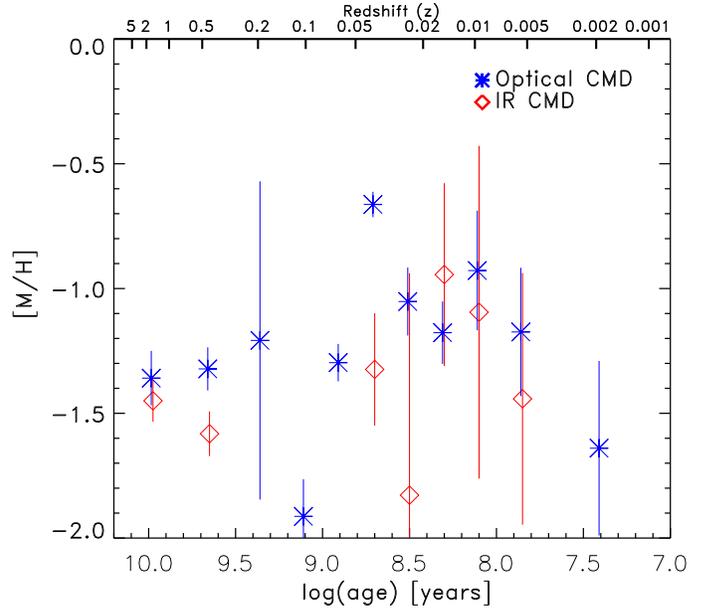}
\caption{\label{fig:met} The MATCH derived metallicity of new stars at each epoch from  the optical (blue asterix) and IR (red diamonds) CMDs.  Both show a preference for metal poor stars from roughly 1/30th to 1/10th  solar.  There may be a slight increase in metallicity over time.  For timebins with negligible star formation estimated in the IR CMDs, no metallicity is plotted. }
\end{figure}

\subsubsection{Metallicity}
Figure \ref{fig:met} shows the MATCH measured metallicity of new stars  at every epoch.  Again results from both the optical (blue) and IR (red) CMDs are shown, however, in time bins where no star formation was measured, no metallicities are shown (several IR time bins).  Both suggest that the chemical history of KKH~98 has been dominated by low metallicity star formation, producing stars with typical $[M/H]\sim -1.5$ to $-1.0$.  There may be a trend to more metal rich stars at more recent times.   MATCH can also be run in a mode where the metallicity is constrained to continually increase, as would be expected in a closed box galaxy model.  When run in this mode, the MATCH results do not change appreciably.

\begin{figure}
\includegraphics[trim=40 0 0 0,scale=0.55]{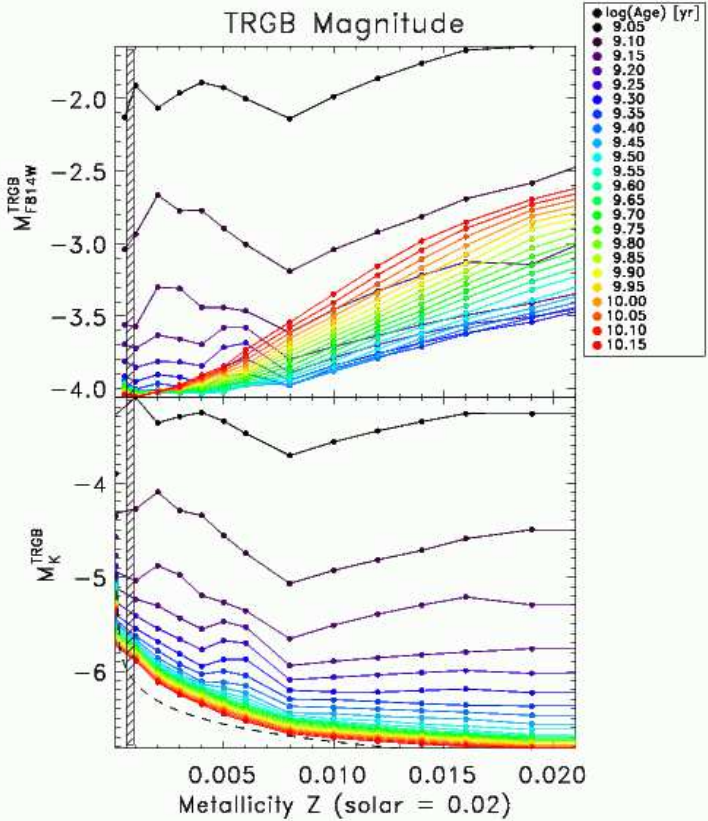}
\caption{\label{fig:trgb2} The magnitude of the tip of the red giant branch (TRGB) as a function of metallicity and age in the F814W (top) and $K$-bands (bottom) as derived by the isochrones of \citet[][colored lines]{Marigo08}, the same isochrones used by MATCH.  Also shown is the metallicity dependence on the $K$-band TRGB as estimated from old Galactic globular clusters \citep[dashed black line from][]{Valenti04}.  The MATCH-derived metallicity of KKH98 of $[M/H]=-1.4$ is indicated by a hatched vertical region.  For metal poor populations, $M_{F814W}^{TRGB} $ is roughly constant with metallicity, and for the metallicity of KKH~98, $M_{F814W}^{TRGB} =-4.0$.  For metal rich systems, the $M_K^{TRGB}$ is roughly constant with metallicity.  At the metallicity of KKH~98, the \citet{Valenti04} relation suggests $M_K^{TRGB}=-6.05$, whereas the \citet{Marigo08} model isochrones suggest a $K$-band tip about 0.2 mags fainter. The latter estimate gives a more consistent distance to KKH~98. }
\end{figure} 

\subsubsection{Reddening and Distance}

MATCH  provides an estimate of the  foreground dust obscuration and distance to KKH~98.  By varying these two parameters, MATCH effectively normalizes the models in color magnitude space so that they best match the observational data.  Because different stellar evolution models will have different normalizations, the best fit reddening and distances reported by MATCH are model-dependent and therefore should not be considered measurements of the actual distance and foreground reddening.  However, a comparison of the reddening and distance results from the two CMDs presented in this paper will  help inform whether consistent results are possible from the optical and IR data sets.

For the optical data, MATCH infers a global $A_V= 0.52 \pm 0.06$.  Using the \citet{Cardelli89} reddening law, this translates to an $A_I=0.23$.  To within the uncertainties, the IR data gives a  similar dust absorption of $A_V=0.42\pm0.08$.  Using the optical CMD, MATCH finds a distance modulus of $DM=27.07\pm0.09$.  The IR CMD produces a similar distance modulus of $DM=27.10\pm0.07$.  To within the uncertainties, both estimates are similar to distances quoted in the literature --- e.g. $DM=26.95\pm0.11$\citep{Karachentsev02}  and $DM=27.023$ \citep{Dalcanton09}.  More on the distance to KKH~98 is given below.  

\subsection{Estimating Distance from the Tip of the Red Giant Branch  \label{sec:TRGB}}
While the MATCH-derived distances arise from normalizing the entire CMD to models, specific features in the CMD can also be used to estimate the distance.  One of the best distance indicators available in the KKH~98 data set is the F814W-band magnitude of TRGB.  Figure \ref{fig:trgb2} shows the F814W-band magnitude  of the TRGB as a function of metallicity and age, using isochrones from \citet{Marigo08}\footnote{Models from CMD 2.1 http://stev.oapd.inaf.it/cmd}.  For old, metal-poor stellar populations the TRGB has an absolute magnitude of $M_{F814W}^{TRGB}=-4.0$ (Figure \ref{fig:trgb2}). We estimate the apparent magnitude of the TRGB in KKH~98 to be at F814W$=23.2 \pm 0.1$.    Correcting for reddening using $A_I=0.23$, this gives a distance modulus (DM) of $DM=26.97 \pm 0.10$, in good agreement with \citet{Karachentsev02} who found $DM=26.95\pm0.11$ from the $I$-band TRGB.

The apparent $K$-band magnitude of the TRGB ($K=21.2\pm0.1$ for KKH~98) can also be used to estimate the distance of KKH~98.   In Galactic globular clusters with old stars, the $K$-band magnitude of the TRGB has been  shown to be a function of metallicity \citep{Valenti04}.  \citet{Gullieuszik07} parameterize this empirical function as:

\begin{equation}
\label{eqn:met}
M_K^{TRGB}=-6.92-0.62\;[M/H].
\end{equation}

\noindent
Assuming the MATCH-estimated KKH~98 metallicity for old stars of $[M/H]=-1.4$, equation \ref{eqn:met} gives $M_K^{TRGB}=-6.05$. Given the apparent TRGB magnitude of $K=21.2\pm0.1$ and negligible dust absorption in the $K$-band, we calculate a KKH~98 distance modulus of $DM=27.25\pm 0.10$.  This distance measurement is somewhat farther than the distance estimated from the apparent F814W-band magnitude of the TRGB.  

However, another estimate of the absolute $K$-band TRGB can be made from the \citet{Marigo08} models shown in Figure \ref{fig:trgb2}, which are also the models used by MATCH. These models suggest that, at low metallicities and old ages, the $K$-band magnitude of the TRGB is 0.2 magnitudes fainter than the value from Equation \ref{eqn:met}.  Using the \citet{Marigo08} models, we find a distance modulus of $DM=27.05\pm 0.10$, which is in very good agreement with the DM derived from optical CMD.

\section{Discussion \label{sec:discuss}}
Using high spatial resolution optical (\HST) and near-IR (Keck AO) images, we have constructed CMDs of the stellar populations within dwarf irregular galaxy, KKH~98.  From these data, we reconstructed the full star formation history of the galaxy, the first time this has been done in the IR for a galaxy beyond the Local Group.  Below we compare the optical and IR results and demonstrate how these techniques may be useful for galaxies at larger distances.


\begin{figure}
\center
\includegraphics[scale=0.4]{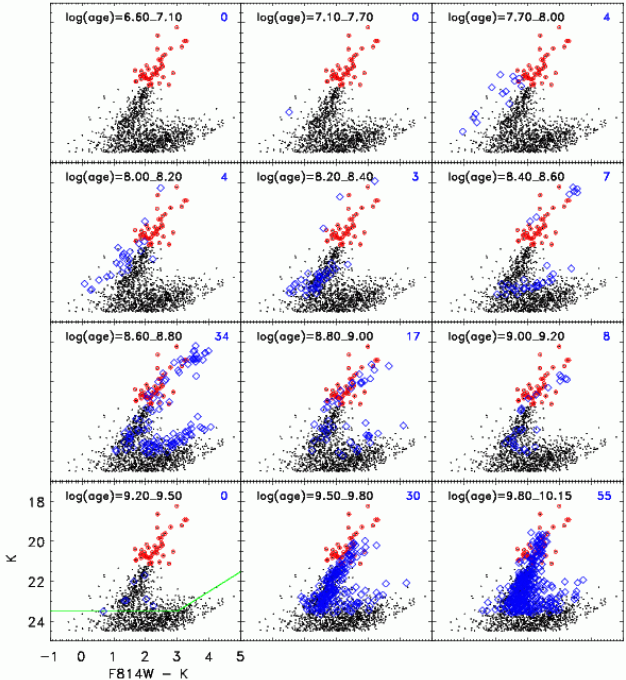}
\caption{\label{fig:ik_fixbi} Same as Figure \ref{fig:ik_fixik}, only now the models (blue diamonds) are created using the optically-derived  SFH.  The over-production of the AGB in the old time bins is basically identical to that for the IR derived star formation history.  However, in this case, significant numbers of AGB stars are also predicted at younger ages. The young model AGB stars are not only over-produced, but especially for log(age) $=$ 8.6 to 8.8, their colors are also typically 0.5 to 1  magnitude redder than the actual AGB populations in KKH~98.  }
\end{figure}

\subsection{Comparison of Optical and IR Results}

Although the number of stars available for analysis in the IR was an order of magnitude smaller than for the deep \HST\ optical data,  the bulk of measured properties --- including the broad trends in star formation history and metal enrichment --- agree between the two data sets. There are some differences however.  The most obvious is that the IR data set is not sensitive to the youngest stars (i.e. age $< 0.1$ Gyr), because the bluest main sequence stars emit the bulk of their radiation in the UV and optical. Therefore, \HST\ optical observations are required to constrain the star formation histories at the youngest ages.  

There are also differences in the details of the star formation histories at ages of $\sim0.1-1$ Gyrs ago,  precisely when AGB stars are expected to dominate the information content of the IR CMD.  For instance, 0.5 Gyrs ago the optical data suggest a burst with a star formation rate four times the prediction of the IR data.  The evidence for this burst comes from the well-populated MS turn-off of the optical CMD at that age (Figure \ref{fig:bi_fixbi}).  

A burst of this age should produce AGB stars observable in our optical and IR CMDs.  Unfortunately,  MATCH predicts that this burst should produce roughly the entire observed AGB population, which would only make sense if there were no AGB production at older times (Figures \ref{fig:bi_fixbi}).  The over-production of AGB stars by the models is even more striking if we model the burst in the IR.  Figure  \ref{fig:ik_fixbi} shows a realization of the model IR CMD created with the optically-derived star formation history.  Not only does the model predict large numbers of AGB stars at 0.5 Gyrs ago, it also suggests that their colors should be $0.5-1$ magnitude redder than the AGB populations seen in the data.  As the SFH derived from the IR data relies heavily on the AGB sequences, it is now clear why the IR data under predicts the star formation in this time bin; there are few AGB stars in the KKH~98 data that match the luminous, red AGB stars predicted by the models for a burst of that age.  Figure \ref{fig:ik_fixbi} also suggests that age/metallicity degeneracies are not producing the discrepancy; the predicted AGB stars are well separated from the observed AGB stars in color space.

Not only do the models over-produce young metal poor AGB populations (and predict redder colors than are seen), they also over-produce old metal poor AGB populations (Figures \ref{fig:bi_fixbi} and \ref{fig:ik_fixik}).  For populations older the 3 Gyrs, the models predict $\sim57 \pm 8$ (optical, assuming Poisson uncertainties) and $\sim79 \pm9$ (IR) AGB stars in KKH~98.  These predictions are both larger than the total number of 42 observed AGB stars, which should presumably include some younger AGB stars as well. Taken together, these results suggest issues with models of AGB stars especially for metal poor populations such as those found in KKH~98.  

The modelling of AGB depends crucially on the prescriptions for two difficult-to-model phenomena: mass loss and the efficiency of third dredge up episodes. To better address the uncertainties related to these processes, modellers resort to synthetic codes in which a few parameters are tuned to fit observational data, typically from the Magellanic Clouds \citep[e.g.][]{Marigo07}. 



While  synthetic AGB codes have successfully reproduced the numbers, luminosities, and colors of AGB stars in the Magellanic Clouds \citep{Cioni99, Nikolaev00, Marigo07}, the situation in other galaxies has been less encouraging.  For instance,   \citet{Gullieuszik08} showed that models calibrated on the LMC and SMC over-predict the AGB lifetimes of the old, metal poor  dwarf spheroidal, Leo~II, by as much as a factor of six.  Girardi et al. (in preparation) reaches a similar conclusion based on a larger sample of ANGST galaxies.

Perhaps more relevant to the present paper are the conclusions from a similar work from \citet{Held10}: In Leo I dSph, a metal poor galaxy containing young stellar populations (therefore more similar to KKH~98), they find that the carbon stars predicted by the models are about twice those observed. We may be seeing a similar issue in KKH~98, where we would need to roughly triple the AGB population in the data to match the models, and the difficulties with the AGB seem to be particularly problematic at younger ages.  We will potentially be able to better constrain AGB models in this vital time range with observations of the AGB populations of a wide variety of galaxies, as will be possible with the ANGST WFC3 survey.

\subsection{Application of These Techniques in Other Systems}

In KKH~98, the IR bright stars are well separated, with typical separations of $~\sim1\arcsec$.  Assuming these separations are common within other galaxies, crowding should not limit the technique until much larger distances (up to 10 Mpc for Keck).   The primary limiting factor for future ground based observations with AO will therefore be sensitivity above the thermal background.  Our study was completed with a 45 minute exposure.  At larger distances, significantly more exposure time, or better AO correction will be required to reach the depth of our study ($\sim2$ magnitudes below the TRGB).  For instance, to reach comparable S/N in a galaxy twice as distant would require 16 times the total exposure time, or 12 hours.  There are, however, several ways in which this time requirement could be reduced.  

1) Improved AO performance.  The observing conditions during this run were sub-optimal with only $\sim20$\% Strehl ratio (ratio of peak luminosity in the PSF to theoretical maximum peak luminosity for a diffraction limited image).  With the Keck AO upgrade, typical Strehl ratio's of 30\% or more are common.  A higher Strehl ratio will dramatically increase the sensitivity above background.  In addition, if the Strehl ratio were more uniform across the image larger numbers of faint stars would be well photometered towards the edges of the image.    More uniform AO performance will be possible with upcoming Multi-Conjugate AO systems such as the one being developed for Gemini south, which will have an isoplanatic patch 4 times the size of the Keck AO system.  

2) At shorter wavelengths, the thermal IR background is reduced.  In $H$-band, \citet{Melbourne07a} showed that it was possible to photometer an $H=24.0$ [Vega] point source to a photometric precision of 17\% in an hour long Keck AO exposure.  The $H$-band sensitivity is significantly deeper than the $K$-band limit.  The drawbacks for using this strategy are, (1) the AO system typically achieves smaller Strehl ratios at shorter wavelengths, and (2) there is a smaller wavelength lever-arm in the CMD when using $H$-band (1.6 $\mu$m) data compared with $K$-band (2.2 $\mu$m).  Despite these limitations $H$-band may be a good strategy for galaxies at larger distances than KKH~98.

3) With near-IR WFC3 observations from \HST, both the thermal background and  field of view issues can be solved.  We expect the planned WFC3 observations of the ANGST sample to reach significantly deeper than our Keck AO observations.  These observations will be made in both the $H$ and $J$-bands.  The roughly factor of 3 loss in resolution should be relatively unimportant at the distances of the ANGST sample, where we have shown that the IR luminous stars are typically not crowding limited.  In more distant galaxies, crowding will be a problem for \HST at these wavelengths.  

4) Future instruments such as AO on a Thirty Meter Telescope or JWST should also overcome the limitation of our current data set.  These telescopes will potentially resolve individual stars in galaxies out to the Virgo Cluster (18 Mpc) and beyond.

\section{Conclusions} 
We obtained high spatial resolution $K$-band imaging of dwarf irregular galaxy KKH~98 using the Keck LGS AO system. The resolution of the AO images ($\sim0.1\arcsec$) was a good match to the existing \HST\ optical images.  In addition, the AO PSFs had a significantly higher contrast above the background compared with seeing limited images, resulting in 15\% photometry at $K\sim22$ [Vega] for our 45 minute exposure \citep[similar to the Keck AO photometric uncertainty found by][in the Local Group dwarf IC 10]{Vacca07}.   

From these images, we created optical and near-IR CMDs of the stars in KKH~98, and estimated the star formation and metal enrichment history of the galaxy.  On average, KKH~98 has experienced a low level ($\sim5 \times10^{-4} M_{\odot}$ yr$^{-1}$) star formation rate over its lifetime, and has maintained a metallicity of roughly 1/30th --- 1/10th solar.  In addition, there is evidence for a prolonged period  of quiescence (1-3 Gyrs ago) where little star formation occurred.  Despite the relatively small number of IR bright stars (592 compared to 2539 in the optical),  the results from the IR data match those from the optical for the bulk of cosmic time.  

However, there are some discrepancies in the optical and IR derived SFHs for ages $<1$ Gyr.  In the optical CMD, younger populations were primarily constrained by the main sequence turn-offs,  whereas the IR CMD constrained younger populations via the asymptotic giant branch.  Discrepancies between the two can be explained by an over-production of AGB stars in the SFH models.  We estimate that the models over-predict the numbers of AGB stars in KKH~98 by as much as a factor of three.  The over-production of AGB stars in SFH models has been noted previously in studies of low metallicity galaxies \citep{Gullieuszik08}, suggesting that  AGB lifetimes (and mass loss rates) may be metallicity dependent.  

Because the IR bright stars were not crowded, these techniques may be useful for measuring stellar populations at much larger distances, especially with the James Webb Space Telescope and AO on future Thirty-Meter class telescopes. However, improved models of AGB evolution may be required to take full advantage of these techniques in the IR.


\acknowledgments
The adaptive optics data presented herein were obtained at the Keck Observatory, which is operated as a scientific partnership among Caltech, UC, and NASA.  The authors wish to recognize and acknowledge the very significant cultural role and reverence that the summit of Mauna Kea has always had within the indigenous Hawaiian community. The laser guide star adaptive optics system was funded by the W. M. Keck Foundation.     This work has been supported in part by the NSF Science and Technology Center for Adaptive Optics, managed by the University of California (UC) at Santa Cruz under the cooperative agreement No. AST-9876783. S. M. Ammons acknowledges support from the Bachmann instrumentation program through UCO/Lick. L. Girardi was partially supported by contract ASI-INAF I/016/07/0.

\bibliographystyle{apj}
\bibliography{/Users/jmel/bib/bigbib2}
\clearpage

\input{tab1.tex}

\input{tab2.tex}

\end{document}

%% file: tab1.tex
\begin{deluxetable}{lccc}
\tabletypesize{\small}
\tablecaption{Summary of MATCH derived results \label{tab:results}}
\tablehead{\colhead{} & \colhead{Optical CMD}  &\colhead{IR CMD}   }
\startdata
$<$SFR$>$ $M_{\odot}\;yr^{-1}$& $5.18\times 10^{-4}$   & $5.45\times 10^{-4} $ \\
$<M/H>$ & -1.33 &  -1.21 \\ 
Reddening ($A_V$) & $0.52 \pm 0.06 $&  $0.41 \pm0.08$\\ 
Distance &$27.06 \pm 0.09$ & $27.10 \pm0.07$\\
\enddata			 
\end{deluxetable}

%% file: tab2.tex
\begin{deluxetable}{lcccc}
\tabletypesize{\small}
\tablecaption{Stellar density in different evolutionary phases of the CMD for both the KKH~98 data and MATCH models.  \label{tab:cmdnum}}
\tablehead{\colhead{} & \colhead{Optical CMD}  & \colhead{Optical Model}  &\colhead{IR CMD}   & \colhead{IR Model}  }
\startdata
AGB &  42\tablenotemark{a} & 130 & 41\tablenotemark{b} & 92 \\
upper RGB &  851  \tablenotemark{c}  & 800       & 207\tablenotemark{d} & 191 \\
MS &  446\tablenotemark{e} & 421 \\
\enddata	
\tablenotetext{a}{$F814W < 23.2$ and $F475W-F814W > 1.5$}
\tablenotetext{b}{$K < 21.2$}
\tablenotetext{c}{$ 23.2 < F814W < 25.5$ and $1.5 < F475W-F814W < 2.5$}
\tablenotetext{d}{$21.2 < K < 22.5$ and $ 1.0 < F814W-K < 2.5$}
\tablenotetext{e}{$ 27.0 < F814W < 23.0$ and $F475W-F814W < 0.5$}		 
\end{deluxetable}

%% file: ms.bbl
\begin{thebibliography}{47}
\expandafter\ifx\csname natexlab\endcsname\relax\def\natexlab#1{#1}\fi

\bibitem[{{Aaronson} \& {Mould}(1985)}]{Aaronson85}
{Aaronson}, M., \& {Mould}, J. 1985, \apj, 290, 191

\bibitem[{{Aparicio} {et~al.}(2001){Aparicio}, {Carrera}, \&
  {Mart{\'{\i}}nez-Delgado}}]{Aparicio01}
{Aparicio}, A., {Carrera}, R., \& {Mart{\'{\i}}nez-Delgado}, D. 2001, \aj, 122,
  2524

\bibitem[{{Bessell} \& {Brett}(1988)}]{Bessell88}
{Bessell}, M.~S., \& {Brett}, J.~M. 1988, \pasp, 100, 1134

\bibitem[{{Brinchmann} \& {Ellis}(2000)}]{Brinchmann00}
{Brinchmann}, J., \& {Ellis}, R.~S. 2000, \apjl, 536, L77

\bibitem[{{Cardelli} {et~al.}(1989){Cardelli}, {Clayton}, \&
  {Mathis}}]{Cardelli89}
{Cardelli}, J.~A., {Clayton}, G.~C., \& {Mathis}, J.~S. 1989, \apj, 345, 245

\bibitem[{{Carpenter}(2001)}]{Carpenter01}
{Carpenter}, J.~M. 2001, \aj, 121, 2851

\bibitem[{{Cioni} {et~al.}(1999){Cioni}, {Habing}, {Loup}, {Epchtein}, \& {The
  Denis Consortium}}]{Cioni99}
{Cioni}, M.~R., {Habing}, H.~J., {Loup}, C., {Epchtein}, N., \& {The Denis
  Consortium}. 1999, in IAU Symposium, Vol. 190, New Views of the Magellanic
  Clouds, ed. Y.-H. {Chu}, N.~{Suntzeff}, J.~{Hesser}, \& D.~{Bohlender},
  385--+

\bibitem[{{Dalcanton} {et~al.}(2009){Dalcanton}, {Williams}, {Seth}, {Dolphin},
  {Holtzman}, {Rosema}, {Skillman}, {Cole}, {Girardi}, {Gogarten},
  {Karachentsev}, {Olsen}, {Weisz}, {Christensen}, {Freeman}, {Gilbert},
  {Gallart}, {Harris}, {Hodge}, {de Jong}, {Karachentseva}, {Mateo}, {Stetson},
  {Tavarez}, {Zaritsky}, {Governato}, \& {Quinn}}]{Dalcanton09}
{Dalcanton}, J.~J., {Williams}, B.~F., {Seth}, A.~C., {Dolphin}, A.,
  {Holtzman}, J., {Rosema}, K., {Skillman}, E.~D., {Cole}, A., {Girardi}, L.,
  {Gogarten}, S.~M., {Karachentsev}, I.~D., {Olsen}, K., {Weisz}, D.,
  {Christensen}, C., {Freeman}, K., {Gilbert}, K., {Gallart}, C., {Harris}, J.,
  {Hodge}, P., {de Jong}, R.~S., {Karachentseva}, V., {Mateo}, M., {Stetson},
  P.~B., {Tavarez}, M., {Zaritsky}, D., {Governato}, F., \& {Quinn}, T. 2009,
  \apjs, 183, 67

\bibitem[{{Davidge}(2009)}]{Davidge09}
{Davidge}, T.~J. 2009, \apj, 697, 1439

\bibitem[{{Dolphin}(2000)}]{Dolphin00}
{Dolphin}, A.~E. 2000, \pasp, 112, 1383

\bibitem[{{Dolphin}(2002)}]{Dolphin02}
---. 2002, \mnras, 332, 91

\bibitem[{{Dolphin} {et~al.}(2005){Dolphin}, {Weisz}, {Skillman}, \&
  {Holtzman}}]{Dolphin05}
{Dolphin}, A.~E., {Weisz}, D.~R., {Skillman}, E.~D., \& {Holtzman}, J.~A. 2005,
  ArXiv Astrophysics e-prints

\bibitem[{{Frogel} {et~al.}(1990){Frogel}, {Mould}, \& {Blanco}}]{Frogel90}
{Frogel}, J.~A., {Mould}, J., \& {Blanco}, V.~M. 1990, \apj, 352, 96

\bibitem[{{Girardi} {et~al.}(2002){Girardi}, {Bertelli}, {Bressan}, {Chiosi},
  {Groenewegen}, {Marigo}, {Salasnich}, \& {Weiss}}]{Girardi02}
{Girardi}, L., {Bertelli}, G., {Bressan}, A., {Chiosi}, C., {Groenewegen},
  M.~A.~T., {Marigo}, P., {Salasnich}, B., \& {Weiss}, A. 2002, \aap, 391, 195

\bibitem[{{Girardi} {et~al.}(2008){Girardi}, {Dalcanton}, {Williams}, {de
  Jong}, {Gallart}, {Monelli}, {Groenewegen}, {Holtzman}, {Olsen}, {Seth},
  {Weisz}, \& {the ANGST/ANGRRR Collaboration}}]{Girardi08}
{Girardi}, L., {Dalcanton}, J., {Williams}, B., {de Jong}, R., {Gallart}, C.,
  {Monelli}, M., {Groenewegen}, M.~A.~T., {Holtzman}, J.~A., {Olsen}, K.~A.~G.,
  {Seth}, A.~C., {Weisz}, D.~R., \& {the ANGST/ANGRRR Collaboration}. 2008,
  \pasp, 120, 583

\bibitem[{{Gogarten} {et~al.}(2009){Gogarten}, {Dalcanton}, {Williams}, {Seth},
  {Dolphin}, {Weisz}, {Skillman}, {Holtzman}, {Cole}, {Girardi}, {de Jong},
  {Karachentsev}, {Olsen}, \& {Rosema}}]{Gogarten09}
{Gogarten}, S.~M., {Dalcanton}, J.~J., {Williams}, B.~F., {Seth}, A.~C.,
  {Dolphin}, A., {Weisz}, D., {Skillman}, E., {Holtzman}, J., {Cole}, A.,
  {Girardi}, L., {de Jong}, R.~S., {Karachentsev}, I.~D., {Olsen}, K., \&
  {Rosema}, K. 2009, \apj, 691, 115

\bibitem[{{Gullieuszik} {et~al.}(2008){Gullieuszik}, {Held}, {Rizzi},
  {Girardi}, {Marigo}, \& {Momany}}]{Gullieuszik08}
{Gullieuszik}, M., {Held}, E.~V., {Rizzi}, L., {Girardi}, L., {Marigo}, P., \&
  {Momany}, Y. 2008, \mnras, 388, 1185

\bibitem[{{Gullieuszik} {et~al.}(2007){Gullieuszik}, {Held}, {Rizzi},
  {Saviane}, {Momany}, \& {Ortolani}}]{Gullieuszik07}
{Gullieuszik}, M., {Held}, E.~V., {Rizzi}, L., {Saviane}, I., {Momany}, Y., \&
  {Ortolani}, S. 2007, \aap, 467, 1025

\bibitem[{{Heavens} {et~al.}(2004){Heavens}, {Panter}, {Jimenez}, \&
  {Dunlop}}]{Heavens04}
{Heavens}, A., {Panter}, B., {Jimenez}, R., \& {Dunlop}, J. 2004, \nat, 428,
  625

\bibitem[{{Held} {et~al.}(2010){Held}, {Gullieuszik}, {Rizzi}, {Girardi},
  {Marigo}, \& {Saviane}}]{Held10}
{Held}, E.~V., {Gullieuszik}, M., {Rizzi}, L., {Girardi}, L., {Marigo}, P., \&
  {Saviane}, I. 2010, \mnras, arXiv:1001.5412

\bibitem[{{Karachentsev} {et~al.}(2001){Karachentsev}, {Karachentseva}, \&
  {Huchtmeier}}]{Karachentsev01}
{Karachentsev}, I.~D., {Karachentseva}, V.~E., \& {Huchtmeier}, W.~K. 2001,
  \aap, 366, 428

\bibitem[{{Karachentsev} {et~al.}(2002){Karachentsev}, {Sharina}, {Makarov},
  {Dolphin}, {Grebel}, {Geisler}, {Guhathakurta}, {Hodge}, {Karachentseva},
  {Sarajedini}, \& {Seitzer}}]{Karachentsev02}
{Karachentsev}, I.~D., {Sharina}, M.~E., {Makarov}, D.~I., {Dolphin}, A.~E.,
  {Grebel}, E.~K., {Geisler}, D., {Guhathakurta}, P., {Hodge}, P.~W.,
  {Karachentseva}, V.~E., {Sarajedini}, A., \& {Seitzer}, P. 2002, \aap, 389,
  812

\bibitem[{{Loidl} {et~al.}(2001){Loidl}, {Lan{\c c}on}, \&
  {J{\o}rgensen}}]{Loidl01}
{Loidl}, R., {Lan{\c c}on}, A., \& {J{\o}rgensen}, U.~G. 2001, \aap, 371, 1065

\bibitem[{{Maraston} {et~al.}(2006){Maraston}, {Daddi}, {Renzini}, {Cimatti},
  {Dickinson}, {Papovich}, {Pasquali}, \& {Pirzkal}}]{Maraston06}
{Maraston}, C., {Daddi}, E., {Renzini}, A., {Cimatti}, A., {Dickinson}, M.,
  {Papovich}, C., {Pasquali}, A., \& {Pirzkal}, N. 2006, \apj, 652, 85

\bibitem[{{Marigo} \& {Girardi}(2007)}]{Marigo07}
{Marigo}, P., \& {Girardi}, L. 2007, \aap, 469, 239

\bibitem[{{Marigo} {et~al.}(2008){Marigo}, {Girardi}, {Bressan}, {Groenewegen},
  {Silva}, \& {Granato}}]{Marigo08}
{Marigo}, P., {Girardi}, L., {Bressan}, A., {Groenewegen}, M.~A.~T., {Silva},
  L., \& {Granato}, G.~L. 2008, \aap, 482, 883

\bibitem[{{McQuinn} {et~al.}(2009){McQuinn}, {Skillman}, {Cannon}, {Dalcanton},
  {Dolphin}, {Stark}, \& {Weisz}}]{McQuinn09}
{McQuinn}, K.~B.~W., {Skillman}, E.~D., {Cannon}, J.~M., {Dalcanton}, J.~J.,
  {Dolphin}, A., {Stark}, D., \& {Weisz}, D. 2009, ArXiv e-prints

\bibitem[{{Melbourne} {et~al.}(2008){Melbourne}, {Ammons}, {Wright},
  {Metevier}, {Steinbring}, {Max}, {Koo}, {Larkin}, \&
  {Barczys}}]{Melbourne08a}
{Melbourne}, J., {Ammons}, M., {Wright}, S.~A., {Metevier}, A., {Steinbring},
  E., {Max}, C., {Koo}, D.~C., {Larkin}, J.~E., \& {Barczys}, M. 2008, \aj,
  135, 1207

\bibitem[{{Melbourne} {et~al.}(2007){Melbourne}, {Phillips}, {Harker}, {Novak},
  {Koo}, \& {Faber}}]{Melbourne07a}
{Melbourne}, J., {Phillips}, A.~C., {Harker}, J., {Novak}, G., {Koo}, D.~C., \&
  {Faber}, S.~M. 2007, \apj, 660, 81

\bibitem[{{Melbourne} {et~al.}(2005){Melbourne}, {Wright}, {Barczys},
  {Bouchez}, {Chin}, {van Dam}, {Hartman}, {Johansson}, {Koo}, {Lafon},
  {Larkin}, {Le Mignant}, {Lotz}, {Max}, {Pennington}, {Stomski}, {Summers}, \&
  {Wizinowich}}]{Melbourne05a}
{Melbourne}, J., {Wright}, S.~A., {Barczys}, M., {Bouchez}, A.~H., {Chin}, J.,
  {van Dam}, M.~A., {Hartman}, S., {Johansson}, E., {Koo}, D.~C., {Lafon}, R.,
  {Larkin}, J., {Le Mignant}, D., {Lotz}, J., {Max}, C.~E., {Pennington},
  D.~M., {Stomski}, P.~J., {Summers}, D., \& {Wizinowich}, P.~L. 2005, \apjl,
  625, L27

\bibitem[{{Nikolaev} \& {Weinberg}(2000)}]{Nikolaev00}
{Nikolaev}, S., \& {Weinberg}, M.~D. 2000, \apj, 542, 804

\bibitem[{{Noeske} {et~al.}(2007){Noeske}, {Faber}, {Weiner}, {Koo}, {Primack},
  {Dekel}, {Papovich}, {Conselice}, {Le Floc'h}, {Rieke}, {Coil}, {Lotz},
  {Somerville}, \& {Bundy}}]{Noeske07}
{Noeske}, K.~G., {Faber}, S.~M., {Weiner}, B.~J., {Koo}, D.~C., {Primack},
  J.~R., {Dekel}, A., {Papovich}, C., {Conselice}, C.~J., {Le Floc'h}, E.,
  {Rieke}, G.~H., {Coil}, A.~L., {Lotz}, J.~M., {Somerville}, R.~S., \&
  {Bundy}, K. 2007, \apjl, 660, L47

\bibitem[{{Olsen} {et~al.}(2006){Olsen}, {Blum}, {Stephens}, {Davidge},
  {Massey}, {Strom}, \& {Rigaut}}]{Olsen06}
{Olsen}, K.~A.~G., {Blum}, R.~D., {Stephens}, A.~W., {Davidge}, T.~J.,
  {Massey}, P., {Strom}, S.~E., \& {Rigaut}, F. 2006, \aj, 132, 271

\bibitem[{{Rejkuba} {et~al.}(2009){Rejkuba}, {Mouhcine}, \&
  {Ibata}}]{Rejkuba09}
{Rejkuba}, M., {Mouhcine}, M., \& {Ibata}, R. 2009, \mnras, 396, 1231

\bibitem[{{Sarajedini} \& {Jablonka}(2005)}]{Sarajedini05}
{Sarajedini}, A., \& {Jablonka}, P. 2005, \aj, 130, 1627

\bibitem[{{Sawicki}(2002)}]{Sawicki02}
{Sawicki}, M. 2002, \aj, 124, 3050

\bibitem[{{Schlegel} {et~al.}(1998){Schlegel}, {Finkbeiner}, \&
  {Davis}}]{Schlegel98}
{Schlegel}, D.~J., {Finkbeiner}, D.~P., \& {Davis}, M. 1998, \apj, 500, 525

\bibitem[{{Skrutskie} {et~al.}(2006){Skrutskie}, {Cutri}, {Stiening},
  {Weinberg}, {Schneider}, {Carpenter}, {Beichman}, {Capps}, {Chester},
  {Elias}, {Huchra}, {Liebert}, {Lonsdale}, {Monet}, {Price}, {Seitzer},
  {Jarrett}, {Kirkpatrick}, {Gizis}, {Howard}, {Evans}, {Fowler}, {Fullmer},
  {Hurt}, {Light}, {Kopan}, {Marsh}, {McCallon}, {Tam}, {Van Dyk}, \&
  {Wheelock}}]{Skrutskie06}
{Skrutskie}, M.~F., {Cutri}, R.~M., {Stiening}, R., {Weinberg}, M.~D.,
  {Schneider}, S., {Carpenter}, J.~M., {Beichman}, C., {Capps}, R., {Chester},
  T., {Elias}, J., {Huchra}, J., {Liebert}, J., {Lonsdale}, C., {Monet}, D.~G.,
  {Price}, S., {Seitzer}, P., {Jarrett}, T., {Kirkpatrick}, J.~D., {Gizis},
  J.~E., {Howard}, E., {Evans}, T., {Fowler}, J., {Fullmer}, L., {Hurt}, R.,
  {Light}, R., {Kopan}, E.~L., {Marsh}, K.~A., {McCallon}, H.~L., {Tam}, R.,
  {Van Dyk}, S., \& {Wheelock}, S. 2006, \aj, 131, 1163

\bibitem[{{Steinbring} {et~al.}(2008){Steinbring}, {Melbourne}, {Metevier},
  {Koo}, {Chun}, {Simard}, {Larkin}, \& {Max}}]{Steinbring08}
{Steinbring}, E., {Melbourne}, J., {Metevier}, A.~J., {Koo}, D.~C., {Chun},
  M.~R., {Simard}, L., {Larkin}, J.~E., \& {Max}, C.~E. 2008, \aj, 136, 1523

\bibitem[{{Vacca} {et~al.}(2007){Vacca}, {Sheehy}, \& {Graham}}]{Vacca07}
{Vacca}, W.~D., {Sheehy}, C.~D., \& {Graham}, J.~R. 2007, \apj, 662, 272

\bibitem[{{Valenti} {et~al.}(2004){Valenti}, {Ferraro}, \&
  {Origlia}}]{Valenti04}
{Valenti}, E., {Ferraro}, F.~R., \& {Origlia}, L. 2004, \mnras, 354, 815

\bibitem[{{Wainscoat} \& {Cowie}(1992)}]{Wainscoat92}
{Wainscoat}, R.~J., \& {Cowie}, L.~L. 1992, \aj, 103, 332

\bibitem[{{Weisz} {et~al.}(2008){Weisz}, {Skillman}, {Cannon}, {Dolphin},
  {Kennicutt}, {Lee}, \& {Walter}}]{Weisz08}
{Weisz}, D.~R., {Skillman}, E.~D., {Cannon}, J.~M., {Dolphin}, A.~E.,
  {Kennicutt}, Jr., R.~C., {Lee}, J., \& {Walter}, F. 2008, \apj, 689, 160

\bibitem[{{Whitelock} {et~al.}(2006){Whitelock}, {Feast}, {Marang}, \&
  {Groenewegen}}]{Whitelock06}
{Whitelock}, P.~A., {Feast}, M.~W., {Marang}, F., \& {Groenewegen}, M.~A.~T.
  2006, \mnras, 369, 751

\bibitem[{{Williams} {et~al.}(2007){Williams}, {Ciardullo}, {Durrell},
  {Vinciguerra}, {Feldmeier}, {Jacoby}, {Sigurdsson}, {von Hippel}, {Ferguson},
  {Tanvir}, {Arnaboldi}, {Gerhard}, {Aguerri}, \& {Freeman}}]{Williams07}
{Williams}, B.~F., {Ciardullo}, R., {Durrell}, P.~R., {Vinciguerra}, M.,
  {Feldmeier}, J.~J., {Jacoby}, G.~H., {Sigurdsson}, S., {von Hippel}, T.,
  {Ferguson}, H.~C., {Tanvir}, N.~R., {Arnaboldi}, M., {Gerhard}, O.,
  {Aguerri}, J.~A.~L., \& {Freeman}, K. 2007, \apj, 656, 756

\bibitem[{{Williams} {et~al.}(2009{\natexlab{a}}){Williams}, {Dalcanton},
  {Dolphin}, {Holtzman}, \& {Sarajedini}}]{Williams09b}
{Williams}, B.~F., {Dalcanton}, J.~J., {Dolphin}, A.~E., {Holtzman}, J., \&
  {Sarajedini}, A. 2009{\natexlab{a}}, \apjl, 695, L15

\bibitem[{{Williams} {et~al.}(2009{\natexlab{b}}){Williams}, {Dalcanton},
  {Seth}, {Weisz}, {Dolphin}, {Skillman}, {Harris}, {Holtzman}, {Girardi}, {de
  Jong}, {Olsen}, {Cole}, {Gallart}, {Gogarten}, {Hidalgo}, {Mateo}, {Rosema},
  {Stetson}, \& {Quinn}}]{Williams09a}
{Williams}, B.~F., {Dalcanton}, J.~J., {Seth}, A.~C., {Weisz}, D., {Dolphin},
  A., {Skillman}, E., {Harris}, J., {Holtzman}, J., {Girardi}, L., {de Jong},
  R.~S., {Olsen}, K., {Cole}, A., {Gallart}, C., {Gogarten}, S.~M., {Hidalgo},
  S.~L., {Mateo}, M., {Rosema}, K., {Stetson}, P.~B., \& {Quinn}, T.
  2009{\natexlab{b}}, \aj, 137, 419

\end{thebibliography}
